\documentclass[%
 aip,
 cha,
 amsmath,amssymb,
 reprint,%
]{revtex4-2}

\usepackage{graphicx}
\usepackage{overpic}
\usepackage{dcolumn}
\usepackage{bm}

\usepackage[utf8]{inputenc}
\usepackage[T1]{fontenc}
\usepackage{mathptmx}
\usepackage{etoolbox}
\usepackage{xcolor}
\usepackage{hyperref}

\newcommand{\xx}{\mathbf{x}}
\newcommand{\rr}{\mathbf{r}}
\newcommand{\vv}{\mathbf{v}}

\newcommand{\BB}{\mathbf{B}}
\newcommand{\RR}{\mathbb{R}}
\newcommand{\SL}{\mathrm{SL}_2(\mathbb{R})}
\newcommand{\Tr}{\mathrm{Tr}}
\newcommand{\Det}{\mathrm{Det}}
\newcommand{\MM}{\mathsf{M}}
\newcommand{\pyoc}{\texttt{pyoculus}}
\newcommand{\CC}{\mathcal{C}}

\newcommand{\AAA}{\mathbf{A}}

\graphicspath{{fig/}}

\makeatletter
\def\@email#1#2{%
 \endgroup
 \patchcmd{\titleblock@produce}
  {\frontmatter@RRAPformat}
  {\frontmatter@RRAPformat{\produce@RRAP{*#1\href{mailto:#2}{#2}}}\frontmatter@RRAPformat}
  {}{}
}%
\makeatother
\begin{document}

\preprint{AIP/123-QED}

\title[Turnstile area as a measure for chaotic transport in magnetic confinement fusion devices]{Turnstile area as a measure for chaotic transport in magnetic confinement fusion devices}
\author{C. B. Smiet}
 \email{christopher.smiet@epfl.ch.}
\author{L. Rais}%
\author{J. Loizu}
\affiliation{
Ecole Polytechnique Fédérale de Lausanne (EPFL), Swiss Plasma Center (SPC), CH-1015 Lausanne, Switzerland
}%
\author{R. Davies}
\affiliation{Max Planck Institute for Plasma Physics, Wendelsteinstraße 1, 17491 Greifswald}

\date{\today}

\begin{abstract}
  We analyze a measure for the degree of chaos in the magnetic fields of magnetic confinement fusion reactors by calculating the lobe areas of turnstiles - a method developed for characterizing transport into and out of resonance zones in Hamiltonian dynamical systems.
  We develop an efficient algorithm based on an action principle to calculate this quantity directly from the magnetic field, including stellarator magnetic fields which are sourced by a complicated set of three-dimensional coils.
  In the analyzed devices, the turnstile area on the inboard (plasma-facing) manifolds is smaller than the turnstile area on the outboard (wall-facing) manifolds. This indicates that the relative importance of chaos increases as the plasma facing components are moved further away from the plasma. 
  The application of the turnstile area calculation for the design of future reactors will be discussed.
\end{abstract}

\maketitle

\begin{quotation}
Magnetic confinement fusion reactors confine the charged particles of a plasma by channeling them along the twisting field lines of a magnetic field.
A magnetic confinement fusion reactor is designed such that in the center, magnetic field lines lie neatly on nested toroidal surfaces; providing good confinement, but at the edge of the plasma the field is designed to guide the plasma to specific locations on the chamber wall.
  Transport of plasma particles and heat to the wall is largely determined by the amount of field lines (amount of magnetic flux) connecting the region around the plasma to the region near the wall.
A quantity called the 'turnstile area', which is zero if a field is not chaotic and increases as the chaos increases, has been defined in the study of Hamiltonian dynamical systems to quantify the amount of trajectories that connect one region of phase space to another. 
In this paper we use the correspondence between magnetic fields and Hamiltonian dynamical systems, and apply this method to the analysis of the magnetic field in the edge of fusion reactors.
We demonstrate this method on two-dimensional fast-evaluating maps, on a toy model of a tokamak, and calculate this measure of transport in several stellarator configurations, including configurations of the existing device W7-X.
Calculating this transport is important for understanding current and planned fusion reactors, and can be potentially used to control the field structure in the edge as we design future devices.
\end{quotation}

\section{\label{sec:intro} Introduction}
Magnetic confinement fusion has the goal to generate energy through the fusion isotopes of hydrogen, by heating a plasma confined in a strong magnetic field created by a set of coils.
The coils are arranged around a toroidal volume, in which field lines lie on nested surfaces (magnetic surfaces).
The charged particles of the plasma are, to lowest order, confined to move along the field lines (integral curves of this field), as their perpendicular motion is constrained through the Lorentz force.
Because of this, the effective transport of heat and particles along the field lines is up to ten orders of magnitude higher~\cite{Goedbloed_Poedts_2004} along the field lines than perpendicular to the field lines, and temperature and pressure are nearly constant on magnetic surfaces.

\begin{figure}[htbp]
  \includegraphics[width=\linewidth]{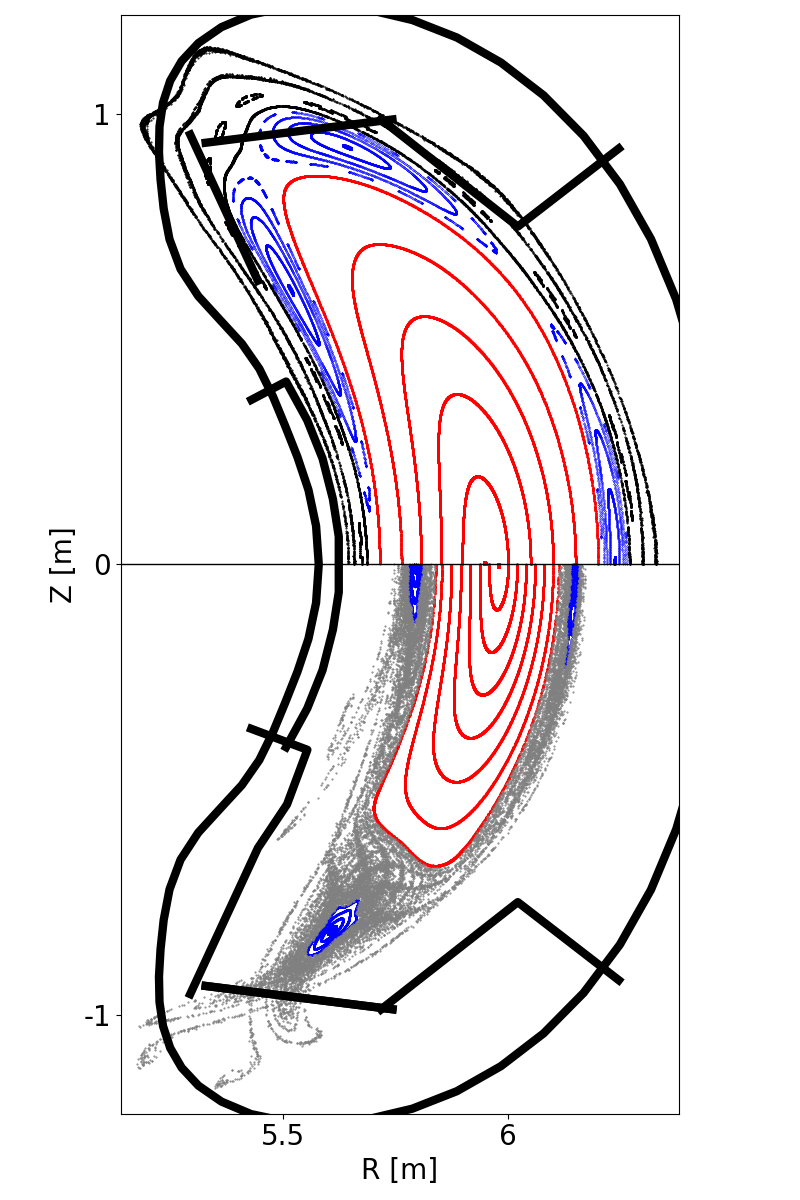}
  \caption{
    Poincar\'e section of the magnetic field in the W7-X stellarator. (top): Standard configuration and (bottom): GYM00+1750 configuration with chaotic field lines.  Field lines in the core are colored red, field lines in a magnetic island surrounding the core are colored blue, field lines outside of the divertor are colored black and field lines that follow chaotic trajectories are colored gray. The plasma facing components and vessel wall are shown in thick black lines.
  \label{fig:introfig} }
\end{figure}

A good way to understand the particle and heat transport in a fusion reactor is therefore to inspect the trajectories of magnetic field lines.
To do this one constructs a Poincar\'e section, by choosing a cross-section of the torus and recording all the points at which a field line intersects that plane.
Such a Poincar\'e section is shown in in figure~\ref{fig:introfig} for the W7-X stellarator.
The trajectories colored in red form the plasma region, where field lines lie on nested magnetic surfaces.
At the very center of this foliation is a single field line, called the \emph{magnetic axis}, named because it is often used as the axis of a toroidal coordinate system.
The magnetic axis is a curve that closes on itself after one toroidal turn (long way around the torus).
The ratio of the number of times a field line winds around the magnetic axis to the number of complete transits of the fusion device is called the \emph{rotational transform} $\imath$.
In tokamaks, which are axisymmetric, it is conventional to use the inverse, called the \emph{safety factor} $q=1/\imath$.

Figure~\ref{fig:introfig} shows two different configurations that can be achieved in W7-X, the top half shows the standard configuration, the bottom half shows the GYM00+1750 configuration achieved by running different currents through the coils.
The trajectories colored in blue form surfaces that do not enclose the magnetic axis, and structures like this are called a \emph{magnetic island}.
These can form on surfaces that have a rational $\imath=m/n$ where $m$ is the number of times field lines on that surface wind around the axis and $n$ the number of full toroidal transits around the device the field line makes to achieve this.
In the standard configuration (top half figure~\ref{fig:introfig}) the island chain in the edge has $\imath=1$ and consists of 5 islands, and is thus an $m/n=5/5$ island chain.
In the GYM00+1750 (bottom panel figure~\ref{fig:introfig}) the rotational transform in the edge is higher and $\imath=5/4$.
At the center of the islands is a field line that closes on itself, and this is called the island \emph{o-point}.
In between the o-points the island regions approach each other, and there is another field line that closes on itself called the island \emph{x-point}.

W7-X has been designed to have this island chain in the edge, to form what is called an (island) \emph{divertor}, which creates spatial separation between the plasma facing components and the plasma itself.
The plasma facing components (PFCs) are also shown in figure~\ref{fig:introfig} with black lines.
Since the field lines in the island connect regions close to the plasma with the outboard side of the island, heat and particles flowing along them can be quickly transported away from the plasma.
Conditions along the divertor path (the path the plasma takes from the core to the PFCs) can be controlled and optimized such that much of the power is dissipated through recombination and radiation, which minimizes damage to the PFCs~\cite{feng2022review}.

When the pressure in the confined plasma is increased, the magnetic field changes, and the field lines around the x-points do not lie on neat surfaces, but become chaotic~\cite{zhou2022equilibrium}. 
There are other configurations in W7-X, generated by running different coil currents, that have chaotic field lines in the edge even when there is no pressure. 
One of these is the GYM00-1750 configuration that will be analyzed in this paper.
A first step towards understanding the transport of heat and particles through such chaotic regions is to calculate the amount of field lines that connect the region close to the plasma to the regions near the PFCs. 

There has also been discussion in the literature about the concept of a \emph{nonresonant divertor}~\cite{boozer2015stellarator, boozer2018simulation}, in which the central region of nested surfaces is not bounded by a clearly identifiable island chain. 
Instead the field transitions from nested surfaces to a region of chaotic field lines. 
Theoretical investigations show that transport through chaotic regions is very important for their operation~\cite{punjabi2014homoclinic}.
Nonresonant divertors are actively being investigated in current experiments~~\cite{garcia2023exploration} and considered for future devices\cite{bader2020advancing}. 

In a tokamak fusion reactor, the main coils that confine the plasma are axisymmetric. 
Axisymmetry precludes the formation of chaotic trajectories, but error fields due to coil misalignment~\cite{piras2010measurement} can introduce some chaos. 
Other tokamak experiments intentionally add non-axisymmetric coils. 
In TEXTOR-DED coils were added to produce an `ergodic divertor`~\cite{schmitz2008identification}. 
The tokamaks Tore-Supra~\cite{ghendrih2009main},  DIIID\cite{evans2005suppression}, KSTAR~\cite{xiao2022investigations} and ITER~\cite{becoulet2008numerical} (among others) have nonaxisymmetric `Resonant Magnetic Perturbation` (RMP)~\cite{evans2015resonant} coils which create a region of chaotic field at the plasma edge.
The transport of heat and particles through the generated chaotic region is actively being studied to predict divertor performance in ITER~\cite{frerichs2020detachment, frerichs2021divertor}. 
RMP fields cause structures called `homoclinic tangles' generated by the field lines which asymptotically limit to the tokamak x-point. 
These can be explicitly calculated in fusion devices~\cite{roeder2003explicit}, and are seen in experiments~\cite{evans2005experimental, jakubowski2007observation}. 

In the case of the GYM+1750 configuration we see a clear example where chaos influences divertor behavior.  To reach the PFCs, the plasma must be somehow transported through a chaotic region. How the heat/particles reach the PFCs (i.e. the structure of the transport, and hence the temperature and density distribution in the edge) is not immediately obvious from the Poincar\'e section.
The same occurs in perturbed tokamaks, and in nonresonant divertors.
Though other transport processes exist, such as neoclassical transport~\cite{beidler2021demonstration} and turbulent transport~\cite{yamada2008anatomy}, parallel transport is likely to be important due to the very high heat conductivity along the field lines.
If we want to quantitatively understand and/or optimize for/against chaos, we need a robust and quantitative toolkit of chaos analysis.

In this paper we will analyze transport through a chaotic region in fusion-relevant configurations using methods lifted from Hamiltonian dynamical systems. 
We will calculate structures similar to~\cite{roeder2003explicit}, and provide an efficient algorithm to calculate the flux enclosed by the structures. 
We will apply these methods to simple maps, to tokamak-like fields and to stellarator fields.

Under suitable conditions~\cite{morrison2000magnetic, kerst1962influence}, the dynamics of magnetic field line flow can be reduced to a $1\tfrac{1}{2}$-dimensional Hamiltonian dynamical system (though this can break down~\cite{duignan2024global, davies2025topological}).
There is a vast body of work studying transport in chaotic systems coming from the Hamiltonian dynamical systems' literature, that can explain a wide-ranging list of physical phenomena. 
The first step in this regard was the frightening observation by Poincar\'e three-body systems such as the earth's orbit around the Sun perturbed by Jupiter are generally chaotic~\cite{poincare1893methodes}. 
These fears were laid to rest by theories by Kolmogorov, Arnol'd~\cite{arnol1963small} and Moser~\cite{moser1973nonlinear}, proving that invariant orbits persist under small perturbations providing barriers to the dynamics. 
The rate of transport through such a barrier when it breaks up can for example explain the ionization probability of atoms in microwave fields~\cite{mackay1988relation}. 
The probability of passing from one region of phase space to another can also be used to calculate probability of ship capsize~\cite{mcrobie1991lobe, naik2017computational}. 

In this paper we focus on a specific quantity that captures the amount of trajectories that pass into and out of a resonance in a dynamical system, called the~\emph{turnstile} (also sometimes called a 'lobe dynamics'). 
This analysis was developed independently by MacKay, Meiss and Percival~\cite{mackay1984transport} and Benimon and Kadanoff~\cite{benimon1984extended}. 
This metric is especially relevant for the magnetic fields in fusion, as the transport into- and out of resonances (the magnetic islands in the edge) directly affects the fusion reactor divertor. 
We analyze this quantity directly in fusion-relevant fields, and develop an algorithm based on an action principle to efficiently calculate this quantity in the magnetic fields of fusion devices. 

Whilst the literature on transport in Hamiltonian dynamical systems is often formulated in terms of differential forms, or otherwise in ways that are agnostic to the dimensionality of the Hamiltonian system under consideration, in this paper we strictly limit ourselves to $1\tfrac{1}{2}$-dimensional Hamiltonian systems and two-dimensional sections.
This will make the exposition simpler to follow for plasma physicists wishing to apply this understanding to the fields in fusion reactors.

This paper is organized as follows: In section~\ref{sec:maps} we first describe three types of maps that have been used to describe fusion-relevant fields: iterated maps, toy models which describe perturbed tokamaks, and the three-dimensional stellarator fields generated by coils.
In section~\ref{sec:fixedpoints} we discuss fixed points of the field line map and we describe an algorithm to find them in any map.
In section~\ref{sec:itmaps} we analyze a specific two-dimensional iterated map, calculate the stable- and unstable \emph{manifolds} of several resonances in it, and illustrate how the turnstile mechanism leads to transport in- and out of a resonance zone. We also describe an algorithm to locate \emph{hetero-} or \emph{homoclinic} orbits in such maps.
In section~\ref{sec:toymodel} we develop a toy model describing an axisymmetric tokamak-like magnetic field, and apply perturbations to it. We describe the calculation of the turnstile area based on an action principle, and use it to calculate how the turnstile area changes with perturbation strength.
In section~\ref{sec:hetero} we apply this to the three-dimensional fields produced by optimized coils of stellarators. We analyze two configurations from the open-source QUASR database~\cite{giuliani2024direct, giuliani2024comprehensive}, that exhibit very different field structures in the edge.
We also analyze a chaotic configuration of the W7-X stellarator called the GYM00+1750 configuration.
Section~\ref{sec:conclusions} concludes with discussion of the competition between chaotic transport in the field and other transport processes, and an outlook for future work.

\section{\label{sec:maps} Iterated maps, toy models, and three-dimensional fields}

An essential tool for studying dynamical systems is the Poincar\'e section; the return map from a section transverse to the dynamics to itself~\cite{poincare1893methodes}.
We can study such maps in three different ways; the simplest is through \emph{iterated maps}, fast-evaluating functions on our space. Second one can use a \emph{toy model}, that describes a Hamiltonian-like function from which a field and perturbations to it are generated. Lastly we will consider fully three-dimensional fields that are generated by coils embedded in space.
In this section we will briefly describe these three types of maps.

\subsection{iterated maps}
It has long been recognized that the magnetic connectivity in fusion devices can be understood through the study of two-dimensional measure-preserving iterated maps, starting with the seminal paper by John Greene~\cite{greene1968two}.
In these maps the difficult and computationally costly task of integrating trajectories (following field lines) is replaced by fast-evaluating analytical expressions of two variables~\cite{abdullaev2004mapping}.

One of the most well-studied and famous iterated maps is the Chirikov-Taylor standard map~\cite{chirikov1971research}, the study of which led Chirikov to formulate the \emph{overlap criterion} determining when the combination of different perturbations leads to chaos.
It has also been used to define the residue criterion~\cite{greene1979method}, a highly accurate method to define the break up of strongly irrational KAM (Kolmogorov, Arnol'd, Moser) surfaces, and to show that after such a surface breaks up, transport through the surface is still hindered by a broken invariant set with the structure of a cantor set, or a \emph{cantorus}~\cite{percival2020variational}.
All these results have had profound impact on the understanding of the magnetic field in fusion reactors.

The standard map is an example of a \emph{twist map}, which equates to fusion magnetic fields where the twist of the field lines of each surface (i.e., the rotational transform $\imath$) is monotonic.
\footnote{more precisely, a line constant in an $S^1$-valued coordinate is mapped to a line that is single-valued in the other coordinate. Equivalently, a map in which each $m/n$ rational orbit (an orbit that traverses the angle coordinate $n$ times for $m$ applications of the map with $n,m\in\mathbb{N}$, $n\neq0$) is either part of one continuous family of $m/n$ orbits, only two singular orbits.}
Aubry~\cite{aubry1983240}-Mather~\cite{mather1982457} theory applies to twist maps and proves (among others) that periodic orbits are related to extrema of an action.
Fusion fields however can and sometimes do violate the twist condition, and these can be studied by \emph{nontwist maps}~\cite{del1996area, morrison2000magnetic}.
These maps also find application in describing particle transport in the fast-particle trajectories~\cite{mugnaine2023nontwist} (a Hamiltonian system that is different from, though related to the one describing field line flow).

In Section~\ref{sec:itmaps} we will analyze an iterated map called the~\emph{tokamap} specifically designed for creating maps that are similar to the field line structure in fusion devices~\cite{balescu1998tokamap}.

\subsection{Perturbed toy models}

A next step in complexity comes when we consider magnetic fields, and the maps that they generate, that are provided by analytical expressions.
There are many different ways to mathematically represent such fields, but a good approach is to start with a formulation that guarantees integrability of the field, and add perturbations to it.

One approach is to exploit the connection to Hamiltonian systems directly, and analytically describe a Hamiltonian.
Through a coordinate transformation to a suitable three-dimensional space we get our magnetic field.
For example, Punjabi and Boozer have analyzed trajectories in Hamiltonian systems that describe a nonresonant divertor~\cite{boozer2015stellarator},  and show chaotic mixing of trajectories just around the last closed surface, with several locations where the trajectories quickly leave the system, i.e. are diverted~\cite{boozer2018simulation, punjabi2022magnetic}.
The locations where the field lines are diverted have been found to be caused by fixed points (see section~\ref{sec:fixedpoints})~\cite{davies2025topological}.

Another method is to describe a simple axisymmetric field, on which we apply perturbations.
In axisymmetry (using cylindrical $(R, \phi, Z)$ coordinates) one can construct a flux function $\Psi(R,Z)$, and a separate function describing the toroidal field.
The poloidal components of the field (defined as the components perpendicular to the $\phi$-direction) are defined by treating $\Psi$ as a stream function.
Although it is easy to identify the Hamiltonian for such fields, this is not often done.
This representation is ubiquitous in tokamak equilibrium reconstruction, where the problem of finding the right flux function can be quickly solved from diagnostic data.

By construction, the field lines lie on the contours of the flux function, and there is no chaos.
We can break the axisymmetry by further adding $\phi$-dependent fields, to have a versatile toy model where the interaction of several symmetry-breaking modes can be studied.

The trajectories are calculated by evaluating this field and integrating curves.
The Poincar\'e map $f(R,Z)$ is calculated by tracing field lines:
\begin{equation}\label{eq:fieldlinemap}
  f^n_{i}(R,Z) = \int_{\phi_0}^{\phi_0 + n 2 \pi}B^i(R, \phi, Z)/B^\phi(R, \phi, Z) \mathrm{d}\phi
\end{equation}
where $i\in \{R, Z\}$, and integration is performed over toroidal angle $\phi$ for a full period, and the superscript $n$ determines how many times around the torus a field line is traced for.

In section~\ref{sec:toymodel} we will develop such a toy model, and analyze the field line flow in it.
We will use the versatility of this model to analyze different perturbations.

\subsection{Maps defined by three-dimensional fields}
In a stellarator, the magnetic field is generated by a complicated set of coils $\{\CC_i\}$, and the magnetic field generated by these is calculated from the Biot-Savart law:
\begin{equation}\label{eq:coilBS}
  \BB(\rr) = \frac{\mu_0}{4\pi}\sum_i \oint_{\CC_i} \frac{j_i \mathrm{d}l \times \rr'}{|\rr'|^3}
\end{equation}
where $j_i$ is the current carried by coil $\CC_i$, $\rr$ the point in three dimensions where the field is evaluated, and $\rr'$ is the vector between $\rr$ and the point on the coil.
Evaluating the Biot-Savart integral is computationally very costly and calculating the field line map can take a significant amount of time per trajectory.

Stellarators are often designed with~\emph{stellarator symmetry}~\cite{dewar1998stellarator} and multiple \emph{field periods} $n_{\rm fp}$.
This means that there are only a few unique coils that are translated and rotated in space, such that the field repeats itself toroidally with period $T= 2\pi/n_{\rm fp} $.
This means that the field line map can be evaluated more efficiently by only integrating over a single period:
\begin{equation}\label{eq:fieldlinemap_symmetry}
  f^n_{i}(R,Z) = \int_{\phi_0}^{\phi_0+nT}B^i/B^\phi \mathrm{d}\phi
\end{equation}
where $T$ is the repetition period in the field.

We note that equation~\eqref{eq:fieldlinemap_symmetry} requires that $B^\phi<0$ everywhere (or $B^\phi >0$).
The fields defined by a finite set of coils, e.g. equation~\eqref{eq:coilBS} do not necessarily fulfill this requirement, and in a given section there can exist field lines that do flow around the torus, but for example get caught close to a coil.
This means that the field line map cannot be defined at every point of $(R, Z) \in (\RR^+ \times \RR)$, but we must restrict the field line map to a region~\cite{davies2025topological}, called the \emph{maximal mappable region}.

In section~\ref{sec:quasr} we will analyze the chaotic fields in several configurations from the open-source QUASR repository~\cite{giuliani2024direct, giuliani2024comprehensive} and analyze the transport in a chaotic configurations of the W7-X fusion reactor.

\section{\label{sec:fixedpoints} The field line map and its fixed points and manifolds.}
In this section we discuss fixed points of the field line map. We demonstrate an algorithm to find them, and discuss the different types of fixed points.
We also define the \emph{stable manifold} and \emph{unstable manifold} that are associated with hyperbolic fixed points.

A fixed point of the field line $n$-map is a point where $f^n(R, Z) = (R, Z)$.
The existence of at least one fixed point is guaranteed by Brouwers' fixed point theorem~\cite{brouwer1911abbildung}, assuming that there is at least some region that maps wholly onto itself.
Given that the goal of a magnetic confinement fusion reactor is to generate a field where field lines foliate nested toroidal surfaces (invariant tori), this is a necessary condition for the fields under consideration.
The `magnetic axis', the field line that lies at the center of this foliation of magnetic surfaces, is a fixed point of the field line map.

\subsection{\label{sec:fixpontalg}locating fixed points}
Fixed points of the map applied $n$ times can be located using a Newton's method.
To first order, using the linearized map around a point $\xx$ the field line map can be approximated as:
\begin{equation}\label{eq:linmap}
  f^n(\xx + \delta \xx)\approx f^n(\xx) + \MM^n \delta\xx
\end{equation}
where the matrix $\MM^n$ (with indices $\MM_{ij} = \partial_j f^n_i$, the labels $i,j$ run over the dimensions $R$ and $Z$) is the Jacobian matrix of the field line map.
$\MM$ is a $2\times2$ matrix with real coefficients.
From this, we can solve for which step $\delta\xx$ the linear approximation brings us to a fixed point, i.e. $f^n(\xx+\delta \xx) = \xx +\delta\xx$:
\begin{align}
  \xx+\delta\xx &= f^n(\xx) + \MM^n\delta\xx \implies \\
  \MM^n \delta\xx - \mathbb{I}\delta\xx &= \xx - f(\xx) .
\end{align}
Thus a series of points is defined by the relation:
\begin{equation}\label{eq:fixpointalg}
  \xx_{k+1} = \xx_k + (\MM^n-\mathbb{I})^{-1}(\xx - f^n(\xx_k)).
\end{equation}
This series rapidly converges in the vicinity of a fixed point, but as any Newton method it can take wild steps far away from fixed points.

For iterated maps (sec 2A), the Jacobian is easily evaluated by differentiating the mapping function, but for integrated maps (sec. 2B and 2C) we must differentiate the result of integrating field lines.
Luckily this can be efficiently calculated by differentiating equation~\eqref{eq:fieldlinemap_symmetry}:
\begin{equation}
  \partial_j f^n_i = \int_{\phi_0}^{\phi_0+nT}\partial_j \frac{B^i}{B^\phi} \mathrm{d}\phi = \int_{\phi_0}^{\phi_0+nT} \frac{\partial_j B^i}{B^\phi} - \frac{B^i}{(B^\phi)^2}\partial_j B^\phi \mathrm{d}\phi
\end{equation}
starting with initial value $\partial_jf^{n=0}_i=\delta_{ij}$ (Kronecker $\delta$), and where the trajectory of integration is determined by the field line itself, equation~\eqref{eq:fieldlinemap_symmetry}.

The above calculations are numerically implemented in the \pyoc ~ package~\cite{pyoculus}, freely available on Github.

\subsection{properties of fixed points}
The Jacobian of the field line map $\MM$ defined in equation~\eqref{eq:linmap} describes the structure of the map around fixed points.
The Poincar\'e map must be measure preserving, which at a fixed point means that $\Det(\MM)=1$.
Hence $\MM\in\SL$, the special linear group of order 2 over the reals~\cite{smiet2019mapping, smiet2020bifurcations}.

The structure of $\SL$ is well known~\cite{lang2012sl2}, and consists of three subsets that act as distinct linear transformations of the plane:
\begin{itemize}
  \item elliptic elements: Points around the fixed point are mapped on trajectories that form invariant ellipses.
  \item parabolic elements: shear mappings, where the point is part of a line of points that are fixed, and points on either side of the line are mapped in opposite directions.
  \item hyperbolic elements: points in the vicinity of the fixed point are mapped along hyperbolic trajectories.
\end{itemize}

Points where $\Tr(\MM)>2$ are hyperbolic elements.
The eigenvalues of the matrix $\MM$  (solutions to the characteristic equation $\lambda_\pm = \left(\Tr(\MM) \pm \sqrt{\Tr(\MM)^2 - 4}\right)/2$  ) are both real and positive, and because $\Det(\MM)=1\implies \lambda_1\lambda_2=1$ we can say without loss of generality $\lambda_1>1$ and $\lambda_2<1$.
The eigenvector corresponding to $\lambda_1$ we call $\vv_1$ or the \emph{unstable eigenvector}, and the eigenvector corresponding to $\lambda_2$ we call $\vv_2$, or the \emph{stable eigenvector}.
By equation~\eqref{eq:linmap} the eigenvectors correspond with directions relative to the fixed point in which points are mapped further away from, respectively closer to the fixed point.
This defines the two asymptotic directions that give nearby trajectories a characteristic x-shape and in fusion literature such points are called x-points (elliptic points conversely are called o-points).

The set of all trajectories that asymptotically approach a hyperbolic point ($\{\xx|\lim_{m\rightarrow\infty}f^{n m}(\xx) = \xx_0\}$) are called the \emph{stable manifold}.
In the linear region around the fixed point (where eq. \eqref{eq:linmap} holds), the manifold is the span of $\vv_2$, as the less-than unity eigenvalue implies that points get mapped closer, but the manifold extends much further as will be discussed in the next section.
The set of trajectories that asymptotically leave the fixed point, or conversely, whose limit $\{\xx|\lim_{m\rightarrow - \infty}f^{n m}(\xx) = \xx_0$ are called the \emph{unstable manifold}, which starts off in the linear regime as the span of $\vv_1$.
Since plasma can follow the field lines in both directions, heat and plasma can be transported both towards and away from the hyperbolic fixed point along both manifolds. 

\section{\label{sec:itmaps} Iterated two-dimensional maps: calculating manifolds. }
In this section we will use two-dimensional iterated maps to illustrate the stable and unstable manifold, and transport through a resonance zone.
We will use the Tokamap, a class of maps developed specifically to describe fusion magnetic fields~\cite{balescu1998tokamap}.
First we will describe the map itself and how it is calculated.
Next we will locate fixed points of this map, and the manifolds associated with them.
Finally we will describe how the intersections of these manifolds determine transport through the system via the `turnstile' mechanism.

The Tokamap has the form:
\begin{align}\label{eq:tokamap}
  f(\psi_{\nu}, \vartheta_{\nu}) = (\psi_{\nu+1}, \vartheta_{\nu+1}) \\
  \psi_{\nu+1} = \tfrac{1}{2}\left[P(\psi_\nu, \vartheta_\nu) + \sqrt{P(\psi_\nu, \vartheta_\nu)^2+ 4 \psi_\nu}\right], \\
  \vartheta_{\nu+1} = \vartheta_\nu + W(\psi_{\nu +1}) - \frac{K}{(2\pi)^2}\frac{1}{(1+\psi_{\nu+1})^2} \cos(2\pi \vartheta_\nu).
\end{align}
Where $K$ is a parameter determining a perturbation strength. $P$ is given by:
\begin{equation}
  P(\psi_{\nu}, \vartheta_\nu) = \psi_\nu - 1 - \frac{K}{2\pi} \sin(2\pi\vartheta_\nu)
\end{equation}
and $W(\psi_{\nu + 1})$ determines the winding of the trajectories (the change in $\vartheta$ at a given $\psi$, which in the unperturbed case corresponds to the rotational transform of the field lines).
We choose a monotonically increasing $W$ given by:
\begin{equation}\label{winding}
  W(\psi) = \tfrac{w_0}{4}(2-\psi)(2-2\psi+\psi^2).
\end{equation}
$W$ has a minimum value of $w_0$ at $\psi_0$.
The parameters $\psi$ and $\vartheta$ are easily identifiable with the variables to describe fusion fields: respectively, the flux label $\psi_p$, which acts as a radial coordinate, and the poloidal angle $\theta$.
Tokamaps have the desirable property that no trajectories pass through the line $\psi=0$, which in fusion devices corresponds to the magnetic axis.

\begin{figure*}[htbp]
  \includegraphics[width=\linewidth]{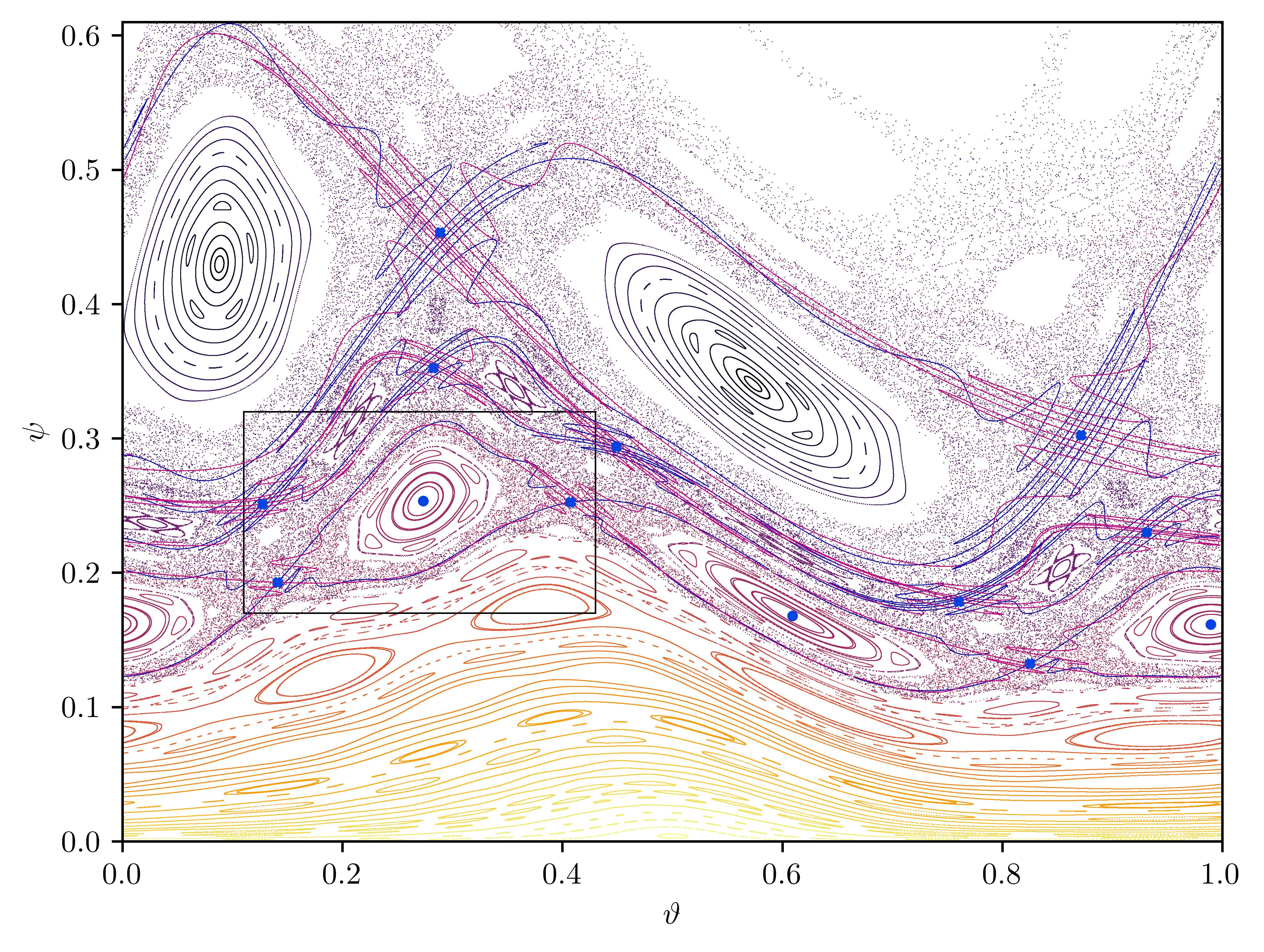}
  \caption{
  Trajectories of a Tokamap configuration, showing an inner region of mostly closed surfaces, bordered by a chaotic region in which transport is determined by field structures surrounding large resonances.
  80 trajectories are calculated for 5000 iterations each.
  The hyperbolic fixed points of the period 3, period 5 and period 2 island are shown in black. 
  The manifolds of these hyperbolic fixed points are shown in blue (for the stable manifolds) and magenta (for the unstable fixed point).
  \label{fig:2dPoincare} }
\end{figure*}

We have plotted trajectories of the Tokamap in figure~\ref{fig:2dPoincare}, with the values $K=4.7$ and $w_0=0.9$.
In the vicinity of $\psi=0$ there is a region where trajectories close in $\vartheta$, corresponding to closed magnetic surfaces.
The trajectories are given a gradient in color starting with yellow at $\psi=0$ and $\vartheta=0.5$ to dark blue at $\psi=0.43$, $\vartheta=0$.
At larger $\psi$, we see trajectories that do not close in $\vartheta$, but form several discrete circles.
These trajectories are analogous to magnetic islands.
The amount of chaotic trajectories in the field increases until a large period 3 island chain is seen, which forms the boundary to a very chaotic field in which a period 5 island chain and a period 2 island chain are seen.

In such a chaotic region, trajectories explore large portions of $\psi$, as is seen from the mixing of different colored trajectories.
This can have the effect of mixing between hot (small $\psi$) and cold (large $\psi$) regions in the plasma which is detrimental to confinement.
Even though there is mixing, we still see the gradient of color from red to blue in the chaotic region, indicating that only a fraction of the red field lines near the plasma will mix in the blue regions further out. 

The amount of field lines that is exchanged between the region on one side of an island chain and the other side is important, because heat and particles are transported along these trajectories. 
Where do the trajectories that pass through the system end up on the wall? 

To analyze this transport we first find the hyperbolic fixed points of each of the island chains.
For the period 3 island chain, we use the Newton's method in equation~\eqref{eq:fixpointalg} applied to the 3-map, with initial guess chosen by eye from the Poincar\'e map.
The three points on the period-3 hyperbolic trajectory are marked with a black 'x' in figure~\ref{fig:2dPoincare}.
Similarly we find and mark the hyperbolic fixed points in the period 5 and period 2 orbits.

We calculate the manifolds of all hyperbolic points.
First we choose one of the $n$ points of the periodic orbit, and evaluate the linearized mapping $\MM^n$ and its eigenvectors.
We choose point a distance of $10^{-5}$ (using the Euclidean norm on $\psi, \vartheta$) from the fixed point along the unstable eigenvector, and evaluate the map of that point.
After confirming that this is in the linear regime, we choose 50 points logarithmically spaced between the initial point and the mapped to point.
We then calculate the trajectories of these 50 distinct points for 12 applications of the $n$-map (12 was found to be a sufficient number to illustrate the manifold).
If $\vv$ is an unstable eigenvector of $\MM$, then $-\vv$ is as well, so we calculate the second unstable manifold as well.
Since the first point of this segment of points maps to the last, the result is a continuous line, that starts straight but becomes more convoluted upon higher iterates of the map.
The result of this calculation are the unstable manifolds of of the x-points, shown in magenta.

We calculate the stable manifolds similarly using the stable eigenvector and the backwards map.
This results in the blue lines show in figure~\ref{fig:2dPoincare}.

\begin{figure*}[htbp]
  \includegraphics[width=\linewidth]{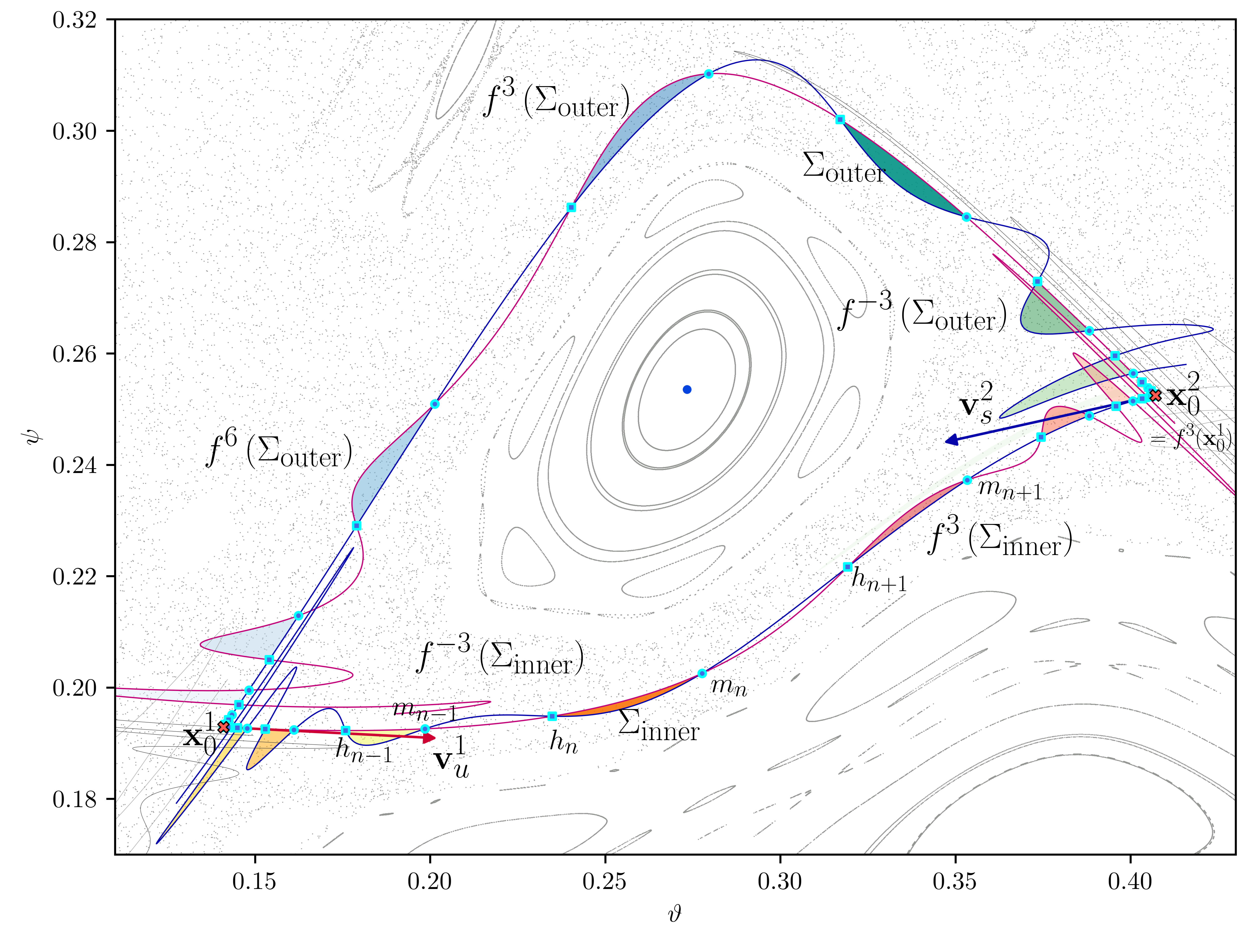}
  \caption{
    Illustration of transport through the barrier created by the first period 3 island chain in the Tokamap.
    Trajectories enter the zone  around the island through a 'turnstile' created by the intersecting manifolds on the inner heteroclinic connection.
    The orange set labeled $\Sigma_\text{inner}$ originates from the plasma region (yellow areas), and is mapped forwards into the region around the island (red areas).
    There is an equivalent set that is mapped from the island zone into the plasma zone.
    The two fixed points from whence the manifolds originate are labeled $\xx_0^1$ and $\xx_0^2$, and their two eigenvectors that are used to calculate the inner heteroclinic orbits are shown by the red (blue) arrows labeled $\vv_u^1$ ($\vv_s^2$) respectively.
    The heteroclinic orbits on the intersections of the manifolds are shown with blue dots. 
    The transport away from the island is governed by another turnstile illustrated with the cyan area, whose backwards map is shown in green, and forwards map in blue.
  \label{fig:zoomin} }
\end{figure*}

In order to understand the chaotic transport of field lines through an island chain, we take a zoomed-in look at a single island indicated by the box in figure~\ref{fig:2dPoincare}.
Figure~\ref{fig:zoomin} shows the two period-3 fixed points labeled $\xx^1_0$ and $\xx^2_0$ and the stable and unstable manifolds associated with them.
The area around the island o-point, and bounded by the manifolds, is called the \emph{island resonance zone} (for the exact definition see~\citep{meiss2015thirty}).
Consider the orange area labeled $\Sigma_\text{inner}$, on the bottom, defined by the intersections of the inner manifolds.
If we apply the forward map $f^{3n}(\Sigma_\text{inner})$, it is mapped to the to the red areas, and when we apply the backwards map, it is mapped to the yellow areas.

It is clear to see that the yellow area starts outside of the resonance zone, but upon several mappings it becomes the red area, which is clearly inside the resonance zone.
There is an identical area that is mapped from inside the resonance zone out.
Because this acts like a `gate' that exchanges a fixed amount of trajectories upon each mapping, it is called the \emph{turnstile} after the gates that limit the flux of persons in and out of buildings.
See~\citet{meiss2015thirty} for a more detailed explanation.

Once a trajectory has entered the resonance zone through the inner turnstile, if it is to pass out of it again, and reach the chaotic region beyond the island chain, it must again pass through a turnstile, this time defined by the intersections of the outer manifolds.
This is indicated with the progression from green via cyan to blue.
If a divertor is placed intersecting the resonance zone, transport from the plasma to this divertor is determined predominantly by the \emph{turnstile area} on the inner connection.

The two manifolds intersect each other along two trajectories which are called the \emph{heteroclinic orbits}.
Since they are in both the unstable manifold of fixed point 1, and the stable manifold of fixed point 2, $\lim_{n\rightarrow\infty} f^n = \xx^1_0$ and $\lim_{n\rightarrow -\infty} f^n = \xx^2_0$.
These orbits must be located in order to define the sections of the manifold that bound a turnstile lobe.

\subsection{\label{sec:clinicfinding} locating homo- and heteroclinic orbits}
Locating the heteroclinic orbits can be done by computing the complete manifolds, and locating their intersections.
Calculating the manifolds is a relatively computationally costly endeavor, as it requires integrating a large number of trajectories for many iterations.
For iterated maps this is not an issue, but if the map is calculated by the numerical integration of trajectories, the computational cost can be significant.
Therefore, we implement an algorithm to find these heteroclinic orbits directly with many fewer required integrations.

We locate a heteroclinc trajectory that starts at $\xx_0^1$, and has its limit at $\xx_0^2$, as illustrated in figure~\ref{fig:zoomin}.
To calculate the heteroclinic trajectory of the inner manifolds, we must integrate along the unstable manifold from $\xx_0^1$, and along the stable of $\xx_0^2$.
To locate this trajectory, we start with a random point in the linear regime along the unstable eigenvector (defined in sec.~\ref{sec:fixedpoints}) of fixed point 1,
\begin{equation}
  \xx_u = \xx_0^1 + \epsilon_u \cdot \vv_u^1
\end{equation}
(here the superscript denotes the fixed point label). We choose $\epsilon_u$ small enough to be in the linear regime, here $10^{-6}$, and we then compute $f^{m n_{\text{map},u}}(\xx_u)$, where $m=3$ is the order of the fixed point, and $n_{\text{map},u}$ is the number of times this initial unstable point is mapped back to the same intersection.

Next we choose a similar point in the stable manifold of fixed point 2:
\begin{equation}
  \xx_s = \xx_0^2 + \epsilon_s \cdot \vv_s^2
\end{equation}
and calculate $f^{-m\cdot n_{\text{map},s}}(\xx_s)$.

If $\xx_s$ and $\xx_u$ are heteroclinic points, then $f^{m\cdot n_{\text{map},s}}(\xx_u) = f^{-m\cdot n_{\text{map},u}}(\xx_u)$, but this is unlikely to be the case for an initial guess.
We therefore minimize the components of:
\begin{equation}\label{eq:clinicfinder}
  \Delta = f^{m\cdot n_{\text{map},u}}(\xx_0^1 + \epsilon_u \cdot \vv_u^1) - f^{-m\cdot n_{\text{map},s}}(\xx_0^2 + \epsilon_s \cdot \vv_s^2).
\end{equation}
This gives a system with two variables ($\epsilon_s$ and $\epsilon_u$) and two constraints ($\Delta_\psi$ and $\Delta_\vartheta$), which is solved using a standard root-finding method \texttt{hybrj} implemented in \texttt{scipy}.

We choose $n_{\text{map},s}=n_{\text{map},u}=8$ for both the stable and the unstable manifold integration, and the values of $\epsilon_{s,u}$ are initially set to $10^{-6}$.
The root-finding converges resulting in a heteroclinic trajectory with 16 points.
There are at least two heteroclinic trajectories (one where the stable manifold intersects the unstable in the positive, the other in the negative). 
To find the other heteroclinic orbit, we perform a new search with $n_{\text{map},s}=8$ for the stable trajectory and $n_{\text{map},u}=7$, and an initial guess for $\epsilon_{s,u}$ halfway between the first and second point of the heteroclinic trajectory that is already found.
This converges to a second heteroclinic trajectory. 
The heteroclinic trajectories thus found are shown in figure~\ref{fig:zoomin} as the blue points.

The turnstiles appear at the intersection between stable and unstable manifolds. In the case above, these manifolds originated from different fixed points, but it is equally possible to perform the calculations on manifolds that originate from the same fixed point.
In that case we are considering an \emph{homiclinic} intersection, and the trajectories are called \emph{homoclinic} trajectories.
The method of locating homoclinic points is identical if $\xx_0^1=\xx_0^2$.

\section{\label{sec:toymodel} Chaotic transport in a perturbed tokamak}
In this section, we will analyze chaotic transport in a tokamak-like field that has been perturbed.
We will first describe the analytical expressions used to calculate an axisymmetric, tokamak-like field and analytical perturbations to it.
We calculate the manifolds in this perturbed system and describe an algorithm to find the homoclinic orbits, the intersections of these manifolds.
Using the homoclinic orbits, we describe an efficient algorithm to calculate the turnstile area using an action principle by MacKay, Meiss and Percival~\cite{mackay1984transport}.

In a three-dimensional field, the fast-evaluating function such as~\eqref{eq:tokamap}, is replaced by an expensive integration of the trajectory as in~\eqref{eq:fieldlinemap}.
Though the variables $(\psi, \vartheta)$ are replaced with $(R,Z)$, all of the discussed algorithms remain the same.
We will start with a axisymmetric field, that resembles the configuration in a tokamak.
Calculation of the manifolds in tokamak geometry has been studied theoretically by~\citet{punjabi2014homoclinic, wei2023invariant}, and their signatures have been shown in experiments~\cite{evans2005experimental}.
Here we build upon this work by implementing our efficient method of calculating the turnstile area, and hence a measure for the degree of chaos and stochastic transport.

We create an analytical toy model that resembles the field in a tokamak fusion reactor.
A general expression for such a field is:
\begin{equation}\label{eq:toytok_axi}
  \BB_\text{circular} = \frac{1}{R} \frac{\partial \psi}{\partial Z}\hat{R} + \frac{1}{R} \frac{\partial \psi}{\partial R}\hat{Z} + B^\phi\hat{\phi}
\end{equation}
Where $\psi$ is a flux surface label, and is related to the $\phi$ component of the vector potential through $\psi = R^2A^\phi$.
We choose $\psi$ such that the field lies on concentric nested circular tori centered on $(R_0, Z_0)$:
\begin{equation}
  \psi(R,Z) = (R-R_0)^2 + (Z-Z_0)^2 = \rho^2
\end{equation}
And define our toroidal field as:
\begin{equation}\label{eq:bphi}
  B^\phi = 2 \sqrt{R^2 - \rho^2}(q_a + \frac{s}{2}\rho^2)/R^2
\end{equation}
which is chosen such as to give a simple expression for the rotational transform $\imath$ or its inverse, $q=1/\imath$ which is more often used in tokamak physics:
\begin{equation}
  \frac{1}{\imath(\rho)} =  q(\rho) = q_a + \frac{s}{2} \rho^2.
\end{equation}
The toroidal field $B^\phi$ (eq.~\eqref{eq:bphi}) is written in terms of a vector potential which is given in the appendix.

In a tokamak, a hyperbolic point is created by a circular separatrix coil underneath or above (or both) the plasma, carrying a current in the same direction as the plasma current.
We calculate this field using the analytical expressions for a circular current loop given in~\cite{simpson2001simple}.
The vector potential for such a loop with radius $R_l$ and positioned horizontally at $Z=Z_l$ is given by:
\begin{equation}\label{eq:loopfield}
  \frac{\mu_0}{4\pi} \frac{4IR_l}{\beta R} \left( \frac{ (2-k^2)K(k^2)-2E(k^2)}{k^2}\right).
\end{equation}
Here $K$ and $E$ are the complete elliptic integral of the first and second kind respectively, $\alpha^2 = (R_l - R)^2 + (Z- Z_l)^2$, $\beta^2 = (R_l+R)^2 + (Z-Z_l)^2$, and $k=1-\alpha^2/\beta^2$.

\begin{figure*}[htbp]
  \begin{overpic}[width=\linewidth]{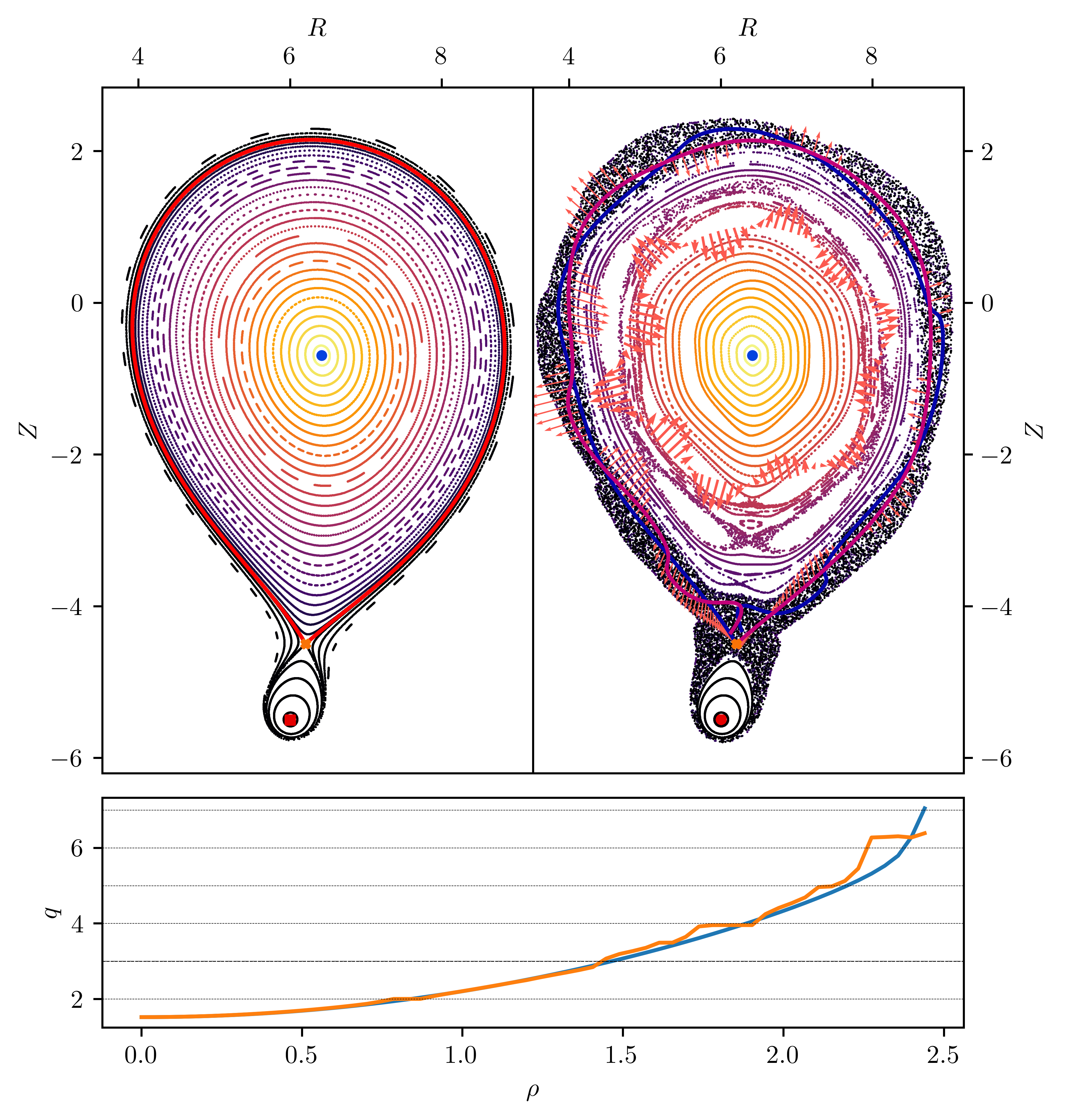}
    \put(15, 90){\textbf{(a)} }
    \put(50, 90){\textbf{(b)} }
    \put(15, 26){\textbf{(c)} }
  \end{overpic}
  \caption{Poincar\'e section of the axisymmetric analytical field and the result of applying a perturbation.
  (a) Axisymmetric field $\BB_{\rm axi}$ with the o-point of the main plasma in blue, the separatrix coil location indicated in red, and the x-point it creates identified by an orange cross. The unstable and stable manifolds that enclose the plasma overlap, and are shown by a magenta and blue line respectively. 
  (b) Effect of applying a $m=6$, $n=1$ perturbation field. The direction the perturbation field is indicated by arrows located on the unperturbed separatrix and on a circle in the plasma region. 
  The perturbation creates islands inside the plasma region, and the formation of chaos around the separatrix. The stable and unstable manifolds (magenta and blue respectively) no longer overlap. 
  (c) safety factor $q$ (inverse of rotational transform $\imath$ measured along a horizontal line from the axis, in both the unperturbed (blue) and perturbed field (orange). The islands in the perturbed field correspond with the regions of flat $q$ at integer values of $q$. 
  \label{fig:toytok} }
\end{figure*}

Equations~\eqref{eq:toytok_axi} and~\eqref{eq:loopfield} are implemented numerically in terms of their respective vector potentials, and the magnetic field is calculated through $\BB = \nabla \times \mathbf{A}$ using automatic differentiation provided by JAX~\cite{jax2018github}.
We choose $R_0=6$, $q_a=1.16$, $s=1.2$ for the axisymmetric field, and for the separatrix coil we choose $R_l=6$, $Z_l=-5.5$, $I=10\pi/\mu_0$.
The resultant field is integrated numerically, and a Poincar\'e section is shown in figure~\ref{fig:toytok} (a).
The fixed points of the 1-map of this field are also located, using an initial guess found manually, and these are the magnetic axis shown with a blue dot, the location of the separatrix coil shown with a red square, and a hyperbolic x-point in between, shown with an orange cross.

Because the fields given by equations~\eqref{eq:toytok_axi} and~\eqref{eq:loopfield} are axisymmetric, the resultant Poincar\'e section in fig.~\ref{fig:toytok} (a) does not exhibit any chaos, and there are no turnstiles.
Now we add a perturbation that depends on the toroidal angle $\phi$ with the form:
\begin{equation}
  A^\phi_{\rm pert} = \mathfrak{f}(\rho) \cos(n\phi + \phi_0) \cos(m\theta +\theta_0)
\end{equation}
where $\phi$ is the usual toroidal angle and $\theta$ is the poloidal angle; $\tan(\theta) = (Z-Z_0)/(R-R_0)$ (with $(R_0, Z_0)$ the location of the magnetic axis, blue circle in fig.~\ref{fig:toytok} (a)).
The radial distribution $\mathfrak{f}(\rho)$ of the perturbation can have any form, we choose one that places the perturbation smoothly everywhere in the configuration by using a Maxwell-Boltzmann probability density function with distribution parameter $d$:
\begin{equation}
  \mathfrak{f}(\rho, d) = \frac{\sqrt{2}}{\sqrt{\pi}}\frac{\rho^2}{d^3}e^{\frac{-\rho^2}{2d^2}}
\end{equation}
From this we calculate
\begin{equation}
  \BB_{\rm pert} = \nabla \times \mathbf{A}_{\rm pert}
\end{equation}

$\BB_{\rm pert}$ lies in the $R, Z$-plane, and is tangent to contours of constant $R^2 A^\phi_{\rm pert}$.
The perturbation field oscillates $m$ times in one $R,Z$-plane, and this pattern rotates $n$ times as the toroidal angle goes from 0 to $2\pi$.
We choose $\phi_0=0$, $\theta_0=0$, distribution width $d=2$, and mode numbers $m=6$ and $n=1$. This is shown in figure~\ref{fig:toytok} (b).

When this perturbation is applied, the field becomes chaotic, as is seen in figure~\ref{fig:toytok} (b).
Islands have opened up on several magnetic surfaces, and the region around the separatrix surface has become chaotic.
The x-point is no longer clearly visible in the Poincar\'e section.
Nevertheless, the fixed point is still there, and it is found using the Newton method described in section~\ref{sec:fixpontalg}, and it is indicated with a orange cross.

In section~\ref{sec:itmaps} we considered a \emph{hetero}clinic intersection, where stable manifolds of one fixed point intersect with the unstable manifolds of a second point in the same orbit, but here, we can consider the simpler where the stable and unstable manifolds originate from the same point; a \emph{homo}clinic connection.
We find the homoclinic trajectory in this integrated map using the same algorithm as described in~\ref{sec:clinicfinding}, and calculate the manifolds in the same way as described in section~\ref{sec:itmaps}.
The results of this calculation are shown in figure~\ref{fig:toytok} (b).

\subsection{\label{sec:tunstile_area} Turnstile area calculation}
We will now describe an efficient algorithm to calculate the turnstile flux $\Sigma_\text{turnstile}$.
One could do this by calculating the shape of the manifold between the two homoclinic orbits, and integrating $\oint \mathbf{A}\cdot \mathrm{d}l$ over this boundary.
This approach is used in numerical solvers such as \texttt{lober}~\cite{naik2017computational}.

When the trajectories are calculated by integrating field lines in three dimensions, resolving the stable and unstable manifolds between two homoclinic orbits can be computationally expensive. 
One would first need to find the homoclinic points (as can be done with the algorithm described in~\ref{sec:clinicfinding}), and then integrate 50-100 trajectories between them. 

Instead, \citet{mackay1984transport} describe a simpler method to calculate the turnstile area. 
For details, we refer the reader to their paper, and the review on turnstiles by Meiss~\cite{meiss2015thirty}. 
Here we give only a brief description, and an intuitive explanation of the method for the specific case of $1\tfrac{1}{2}$D systems such as magnetic fields.  
Their method is based on the Lagrangian form $\lambda$ calculated on homoclinic orbits.
They show that if the map is exact, $\lambda = F(x, x')$ which is one of the standard generating functions for canonical transformations.
They go on to show, for a homo (or hetero)clinic connection, which has two homo (hetero)clinic trajectories, $m_t$ the minimax orbit and $h_t$ the  minimizing orbit, that the homoclinic lobe area can be calculated directly from:
\begin{equation}\label{eq:turnstilesum}
  \Sigma_{\rm turnstile} = \sum_{t=-\infty}^{t=\infty} (\lambda(m_t)-\lambda(h_t))
\end{equation}

In the case of magnetic fields, the Lagrangian form is can be calculated by integrating the vector potential along the field line:
\begin{equation}\label{eq:magneticaction}
  \lambda(m_t) = \int_{\phi=0}^{\phi = nT}\mathbf{A}(\xx)\cdot \mathrm{d}l.
\end{equation}
where the integration is along the field line, it is started at $\xx=m_{t-1}$ and the vector potential is evaluated on each point of the field line as it is evolved simultaneously with the calculation of eq.~\eqref{eq:magneticaction}. 

It seems rather remarkable that the computation of a complicated area in phase space can be performed by the simple summation (equation~\eqref{eq:turnstilesum}) of action integrals (equation~\eqref{eq:magneticaction}).
Though the formulation by~\citet{mackay1984transport} in terms of Lagrangian forms and area forms holds in any dimension, in the case of a three-dimensional flow and a two-dimensional map, this formula has an intuitive explanation.

\begin{figure}[htbp]
  \includegraphics[width=\linewidth]{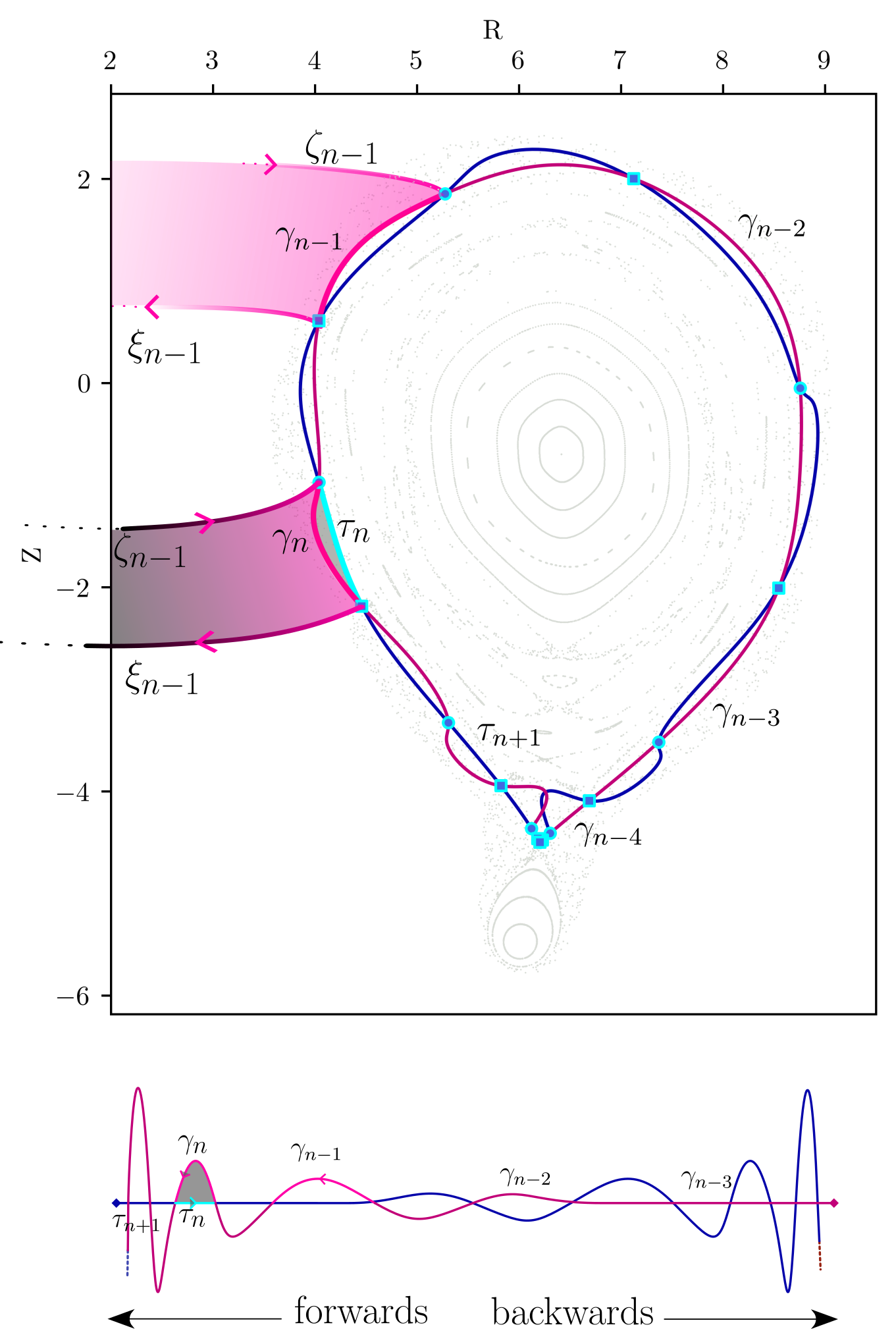}
  \caption{Illustration of the calculation of a turnstile area. The area $\Sigma_n$ shown in gray is bounded by turnstile sections $\gamma_n$ and $\tau_n$ between homoclinic points $h_n$ and $m_n$, is given by the integral of the vector potential over the curve $\tau_n  \cup - \gamma_n$. This integral can be modified to run over $\tau_n \cup \zeta_{n} \cup \gamma_{n-1} \cup \xi_n-1$. This can be repeated until $\gamma_{n-i}$ is in the linear regime of the fixed point. Similarly the integration over $\tau_n$ can be moved forward to $\tau_{n+j}$ in the linear regime of the stable manifold. 
  (bottom): illustration of the turnstile area calculation where the intersecting manifolds have been laid flat. 
  \label{fig:turnstilecartoon} }
\end{figure}

Figure~\ref{fig:turnstilecartoon} illustrates how the calculation of the turnstile lobe can be understood.
To calculate the flux through the dark gray shaded area, which without loss of generality we label the $n$-th occurrence of the lobe, one can use Stokes' theorem and integrate the vector potential $\AAA$ over $\gamma_n$, a section of the unstable manifold, and $\tau_n$, a section of the stable manifold:
\begin{equation}
  \Sigma_{ \rm turnstile} = \int_{\tau_{n}} \mathbf{A}\cdot \mathrm{d}l + \int_{\gamma_n} \mathbf{A}\cdot\mathrm{d}l
\end{equation}
We can now shift the integral over $\gamma_n$ by transporting the entire curve backward along the field lines.  
As this does not change the flux through the area, because the transport is along the field, the result should be identical, and we only need to add the contribution of $\int \AAA\cdot\mathrm{d}l$ along the endpoints, which move along the trajectories $\zeta_{n-1}$ and $\xi_{n-1}$. 
We transport the path $\gamma_n$ all the way back $2 pi$, until it reaches the same section.  
This integral now becomes:
\begin{align}
  \Sigma_{\rm turnstile} &=  \int_{\tau_{n}} \mathbf{A}\cdot\mathrm{d}l + \int_{\xi{n-1}} \mathbf{A}\cdot \mathrm{d}l + \int_{\gamma_{n-1}} \mathbf{A}\cdot \mathrm{d}l - \int_{\zeta_{n-1}} \mathbf{A}\cdot \mathrm{d}l \\
                       &= \int_{\gamma_1} \mathbf{A}\cdot\mathrm{d}l + \lambda(m_n) +\int_{\gamma_1} \mathbf{A}\cdot\mathrm{d}l - \lambda(h_n).
\end{align}
We can repeat this process, shifting the integration over the stable manifold to $\gamma_{n+2}$, $\gamma_{n+3}$, etc, until it is so close to the fixed point that it tends to zero, and all that is left is a sum over the action integrals.

We do the same with the integral over the stable manifold section $\tau_n$, replacing it with action integrals moving it to the section connecting $\tau_{n+1}$, $\tau_{n+2}$ and so forth, until also this contribution shrinks, and we are left with equation~\eqref{eq:magneticaction}.

Instead of integrating the vector potential over two curve segments, whose geometry has to be found by integrating many trajectories, we can instead identify two specific trajectories along which to integrate the vector potential.
One issue in the practical application of equation~\eqref{eq:magneticaction} is when to terminate the sum.
In practice only about 10 applications of the map suffice to bring a point from the linear regime of the unstable manifold into the linear regime of the stable manifold.
Increasing the number of points in the homoclinic trajectories will make the sum converge, but at a large computational cost.
Instead, when we are in the linear regime, we can close the sum by evaluating $\int \mathbf{A}\cdot\mathrm{d}l$ along the straight line connecting the homoclinic points in the linear regime near the fixed point.

\begin{figure*}[htbp]
  \begin{overpic}[width=\linewidth]{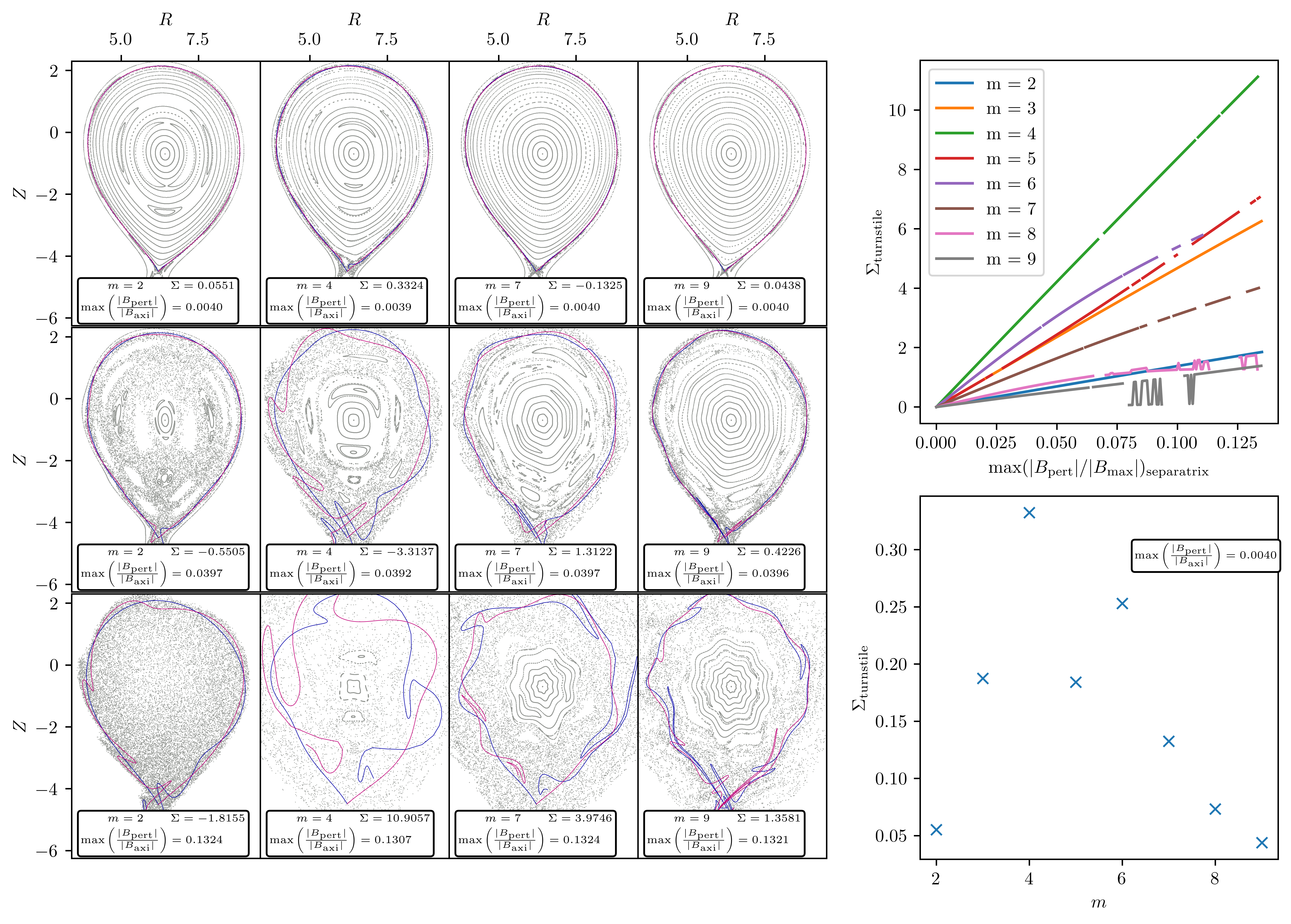}
    \put(4, 65){\textbf{(a)} }
    \put(82, 65){\textbf{(b)} }
    \put(72, 29){\textbf{(c)} }
  \end{overpic}
  \caption{The effect of perturbations on the turnstile area using the same axisymmetric field as in figure~\ref{fig:toytok}. 
  (a): Each column has a different poloidal mode number $m$, and each row has a different amplitude of the perturbation.
  Different mode numbers affect the field and chaotic transport differently.
  Low $m$ mode numbers are resonant with the surfaces inside, and create chaotic fields throughout the volume.
  (b): Turnstile area vs perturbation amplitude. The response to perturbation is mostly linear, but can act nonlinearly at very high amplitudes. 
  Gaps are caused when the calculation fails.
  (c): Turnstile area at fixed perturbation amplitude of $\max\left(\frac{{|B_{{\rm pert}}|}}{{|B_{{\rm axi}}|}} \right)=0.004$. For a given perturbation amplitude the $m=4$ perturbation creates the largest $\Sigma_\text{turnstile}$, followed by the $m=6$ mode.
    \label{fig:perturbationfig} }
\end{figure*}

We now use this algorithm to compare the effects of varying the perturbation mode numbers and perturbation amplitude on the turnstile area in the tokamak toy model equilibrium.
To compare different perturbations, we normalize the perturbation strength the axisymmetric field strength, by evaluating the maximum of $|B_{\rm pert}|/|B_{\rm axi}|$ on the unperturbed separatrix.
Poincar\'e sections at three amplitudes, for the mode numbers $m=2, 4, 7, 9$ ($n$ always 1) are shown in figure~\ref{fig:perturbationfig} (a).
All perturbations create chaotic regions, but the perturbations with lower $m$ create more chaos in the core.
The rotational transform crosses the rational values $m/1$ for those $m<7$ and the mode couples with resonant surfaces there.

The size of the turnstile area as a function of the mode number is shown in figure~\ref{fig:perturbationfig} (b),
The area scales approximately linearly with the applied perturbation, though some modes show nonlinearity at high amplitudes.
Gaps in the graphs are caused by the failure of the algorithm, which occurs sporadically, and is observed to happen when the manifolds are close to intersecting in additional locations, as can be seen close to happening in the bottom row of panel a). 
Despite these few failures, the method is able to calculate turnstile areas over a large range of perturbation amplitudes, and at the maximum applied perturbations the perturbation field is is around 15\% of the unperturbed field strength.
We thus have a reliable method to evaluate the turnstile area over a broad range of perturbation amplitudes, from nearly integrable fields, to moderately chaotic fields up to almost completely chaotic fields.

Different perturbations are more or less effective at creating chaotic transport on the separatrix.
To compare this, we plot the turnstile area at a fixed perturbation amplitude of $\max\left(\tfrac{|B_{\rm pert}|}{|B_{\rm axi}|}\right) =0.0040$ for all different modes.
This is shown in figure~\ref{fig:perturbationfig} (c).

Understanding which modes are most effective at creating chaos in fields is an important and long-standing area of research in tokamak physics, where so-called Resonant Magnetic Perturbation (RMP)~\cite{evans2015resonant} coils are placed around the vessel to create a chaotic edge.
Computing the lobe structures is key to predicting the heat footprints on divertor targets~\cite{joseph2008calculation}.
We expect that the turnstile area calculation could be an effective tool in designing RMP fields.
Furthermore, given the potential risk posed by runaway electrons for future tokamak experiments, there is a big effort in studying mitigation methods, and one proposal are Passive Runaway Electron Mitigation Coil (PREMCs)~\cite{smith2013passive}.
Here a shaped coil is placed near the vessel, and the current induced in it by the abrupt changes in flux brought on by the disruptive events preceding the formation of a dangerous beam of near-relativistic electrons, will induce a current in the coil which creates chaos in the field, disrupting the beam.
Such a coil could be optimized to create maximum chaotic field at desired locations.

\section{\label{sec:hetero} Transport in the Stellarator edge}
We finally turn our attention again to stellarator fields, and in this section we will analyze magnetic island structures occurring at the edge of stellarator configurations. We will first analyze two configurations from the QUASR database~\cite{giuliani2024direct, giuliani2024comprehensive} of stellarator configurations. We will use these configurations to verify the turnstile area calculation. We show that in general, the turnstile areas are larger on the outside (coil-facing) manifolds, and smaller on the inside (plasma-facing) side. Finally we will analyze a specific configuration of the W7-X stellarator called GYM+1750, and show how the structures in the connection length are caused by the manifolds.

Whereas the toy model fields were given by analytical expressions, the fields in stellarators are generated by a set of coils.
The coils are modeled as current-carrying filaments from which the field is calculated using the Biot-Savart integral.
The coils can be represented in different ways, but we use any of representations provided by the \texttt{simsopt}~\cite{landreman2021simsopt} stellarator optimization suite which are based on a Fourier representation to describe the curve.
See the documentation of~\texttt{simsopt} for the implementations.
\texttt{simsopt} also contains optimized routines to evaluate the Biot-Savart integral (equation~\eqref{eq:coilBS}) and vector potential, as well as derivatives thereof.
\pyoc~ has been augmented with an interface to \texttt{simsopt}, such that the field generated by the coils can be used in equation~\eqref{eq:fieldlinemap_symmetry} to define the field line map.

\subsection{\label{sec:quasr} Turnstile area in fields from the QUASR database}
The QUASR database currently contains over 370,000 quasi-axisymmetric stellarator configurations~\cite{giuliani2024direct, giuliani2024comprehensive}.
The entire database can be browsed interactively~\href{https://quasr.flatironinstitute.org/}{https://quasr.flatironinstitute.org/}, and contains configurations with different numbers of coils, field periods and types of symmetry.
Each configuration is defined by a set of coils (and their currents) that produce a quasisymmetric vacuum magnetic field, and a set of toroidal surfaces that were targeted in the optimization. 
The coils can be downloaded in the format used by~\texttt{simsopt}.
Though the agreement between the field and the surfaces was one of the targets in the optimization, this agreement is not always perfect, and many configurations do contain islands, especially if the rotational transform reaches a rational value on the outer surfaces, close to the coils. 

\begin{figure*}[htbp]
  \begin{overpic}[width=\linewidth]{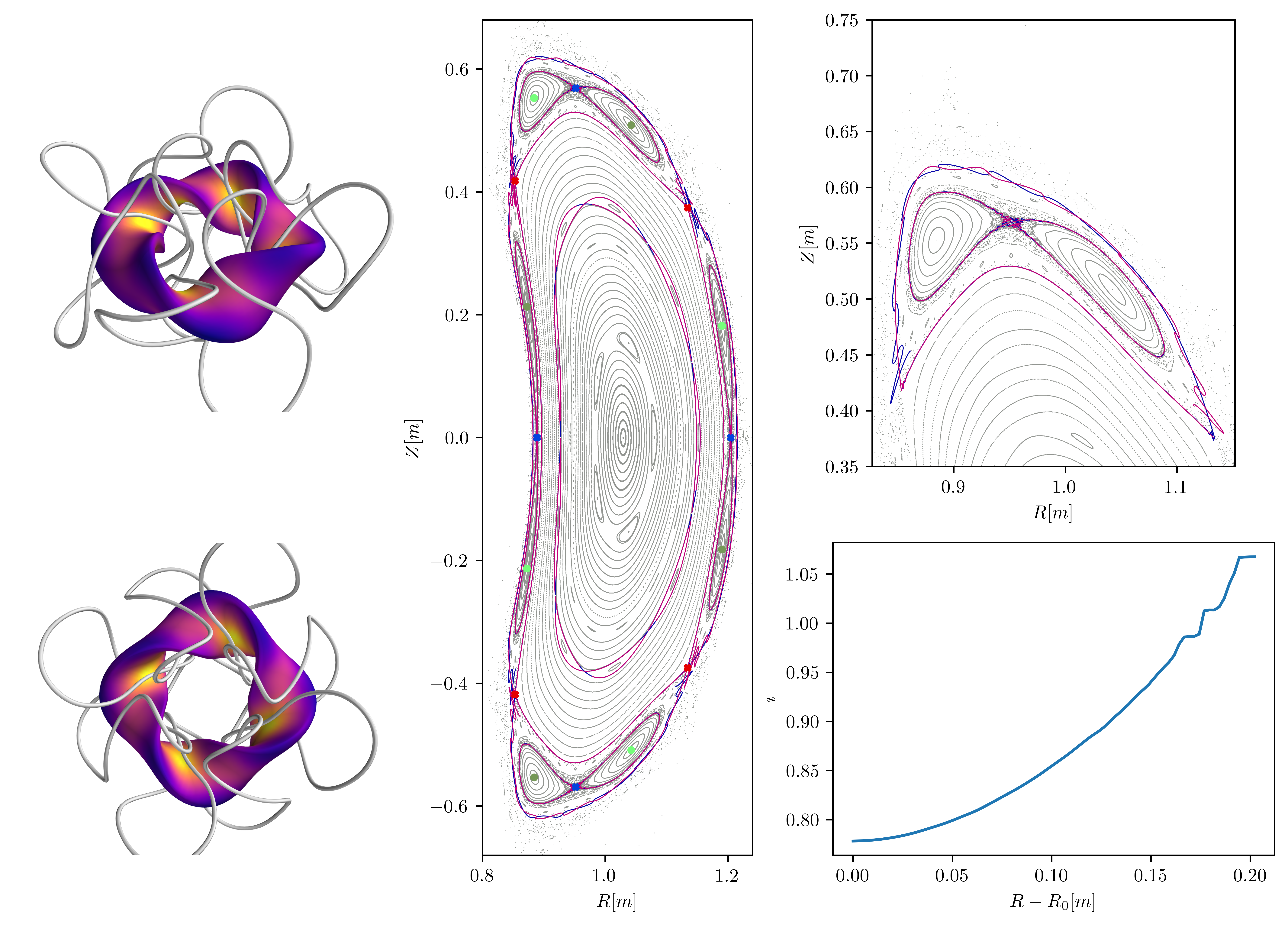}
    \put(3, 65){\textbf{(a)} }
    \put(3, 30){\textbf{(b)} }
    \put(38, 68){\textbf{(c)} }
    \put(69, 68){\textbf{(d)} }
    \put(66, 27){\textbf{(e)} }
  \end{overpic}
  \caption{Analysis of QUASR configuration 242612. 
  (a), (b): Coils and a magnetic surface of the configuration colored by the magnetic field strength.
  (c): Poincar\'e section at $\phi=0$ plane, including the manifolds of the $9/10$ island chain and the outer $4/4$ island chain, as well as manifolds around the hyperbolic point inside of the $4/4$ island chain.
  (d): zoom in of the turnstiles around the edge $4/4$ island. 
  (e): rotational transform profile calculated along the midplane.
  \label{fig:quasr_normal}}
\end{figure*}

We focus on one specific QUASR configuration with identifier 242612, which is a four field period device with only eight coils in total.
The rotational transform starts below 1, and monotonically increases to a value of around 1 at the edge.
We plot the coils and the field strength a surface of the configuration in Figure~\ref{fig:quasr_normal} (a)-(b).
The coils are indicated with gray curves, and a magnetic surface of the configuration is shown, colored by the magnitude of the magnetic field on the surface.

The Poincar\'e section of the field, shown in figure~\ref{fig:quasr_normal} (c), shows a few islands in the core, and a large $4/4$ island chain at the edge, which is surrounded by a chaotic region.
This stellarator configuration thus resembles a conventional stellarator with an island divertor like W7-X, except optimized for quasisymmetry instead of quasi-omnigeneity.
One interesting feature of this configuration is that the resonance in the edge has itself bifurcated, resulting in two elliptic fixed points shown in light and dark green, and a hyperbolic point in between.

We calculate the manifolds, heteroclinic points, and turnstiles for both the inner and outer heteroclinic connections in the 9/10 and 4/4 resonances, as well as the period four homoclinic connections (since here there is no inner or outer, we label these left and right).
The results of this are presented in table~\ref{tab:quasr_normal_table}.

\begin{table}
\begin{tabular}{lcc}
\hline
Manifold & Turnstile Area (in) & Turnstile Area (out) \\
\hline
manifolds 9/10 & $6.3879\cdot10^{-12}$ & $3.7425\cdot10^{-12}$ \\
surrounding heteroclinic manifolds & $9.1613\cdot10^{-07}$ & $3.4471\cdot10^{-05}$ \\
island homoclinic manifolds (left/right) & $1.3626\cdot10^{-06}$ & $1.3627\cdot10^{-06}$ \\
\hline
\end{tabular}

  \caption{Turnstile areas of select resonances in the QUASR 242612 configuration. }\label{tab:quasr_normal_table}
\end{table}

\subsection{\label{sec:areaverification} Turnstile area calculation verification}
\begin{figure}[htbp]
  \includegraphics[width=\linewidth]{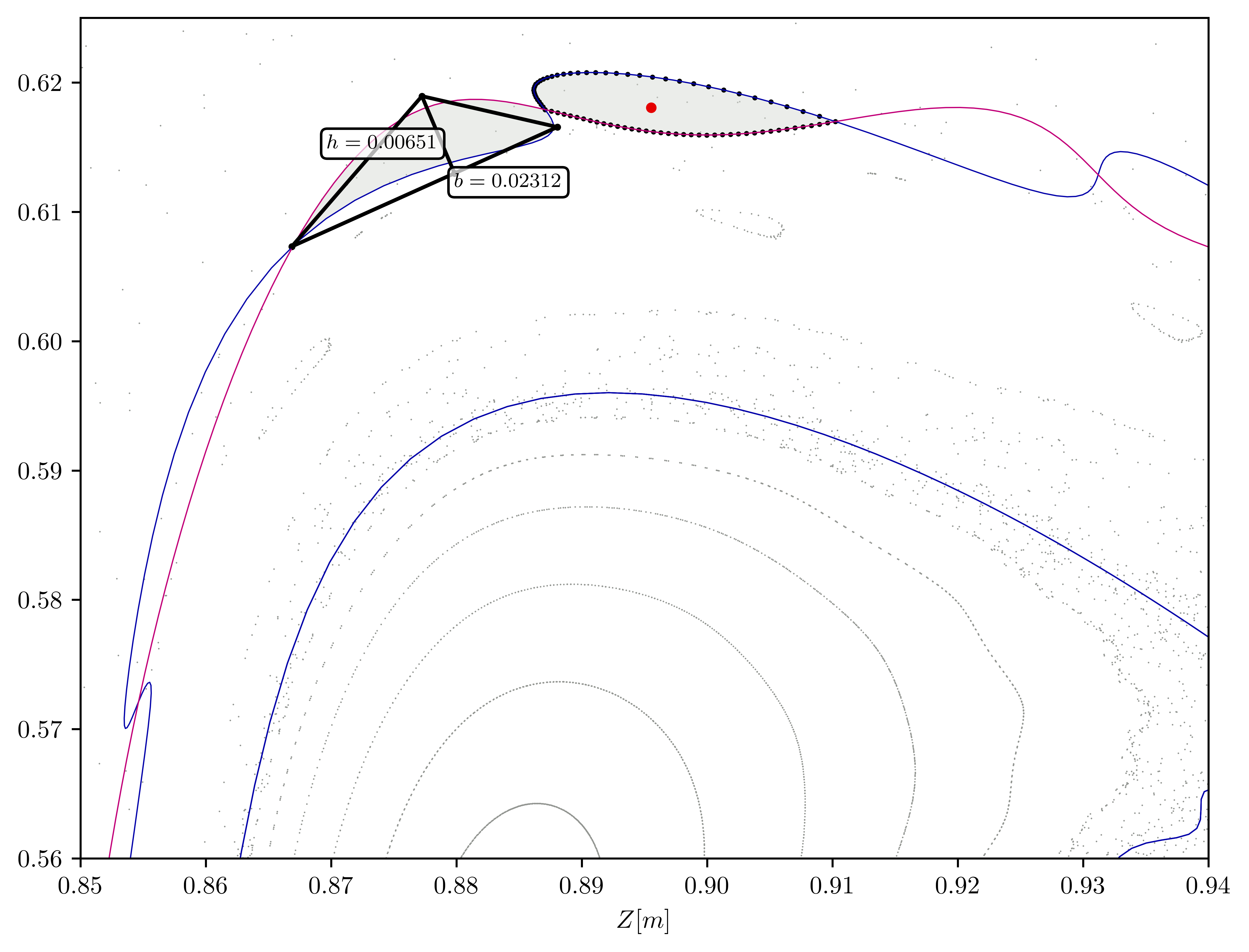}
  \caption{Verification of the action-principle based turnstile area computation against geometrical and direct computation.
  The triangle used for approximating the area is shown in black lines, the points used for calculating the area and integral of the vector potential are shown on the right lobe.
    \label{fig:verification}
  }
\end{figure}

We verify the turnstile area calculation using the field of the QUASR 242612 configuration and the manifold on the outer heteroclinic connection of the 4/4 outermost resonance zone.
First we calculate the area using a simple triangle approximation, by approximating the turnstile lobe with a simple triangle.
The chosen triangle does not exactly match the lobe, but it is chosen such that it roughly agrees.
This very crude approximation is shown with the solid black lines in figure~\ref{fig:verification}, and the triangle has a basis length of $b=0.02312 m$, and a height of $h=0.00651m$ resulting in an area of $7.5264\cdot 10^{-5}m^2$.

We next use a more robust method, where we resolve the turnstile lobe using 40 trajectories between the heteroclinic orbits.
We calculate the area using Gauss's area formula (also known as the shoelace algorithm) which gives the area enclosed by a polygon determined by points ${x_i, y_i}$ by:
\begin{equation}\label{eq:shoelace}
  \Sigma =   \frac{1}{2}  \sum_{i=1}^n y_i \left(x_{i-1} - x_{i+1}\right)
\end{equation}
This method is also used in the code \texttt{lober}~\cite{naik2017computational} to evaluate the area between manifolds.
The result of this calculation is shown in table~\ref{tab:verificationtable}.

\begin{table*}
\begin{tabular}{c|c|c|c|c}
\hline
triangle area $[m^2]$ & Shoelace area $[m^2]$ & $\Sigma_{\rm turnstile}$ $[m^2 \cdot T]$ & $\Sigma_{\rm turnstile}/B^\phi_{\rm center}$ & $\oint_{\rm boundary} \mathbf{A}\cdot \mathrm{d}l$ \\ 
\hline
$7.5264\cdot10^{-05}$ & $7.2437\cdot10^{-05}$ & $3.4471\cdot10^{-05}$ & $7.2436\cdot10^{-05}$ & $3.4444\cdot10^{-05}$ \\ 
\hline
\end{tabular}

  \caption{The turnstile area of the outer heteroclinc connection in the QUASR 242612 stellarator calculated by four different methods. }\label{tab:verificationtable}
\end{table*}

The turnstile calculation returns the amount of flux through the lobe in $[Tm^2]$. In order to get an area, we divide by the field strength in the lobe. We evaluate the toroidal field at the geometrical center of the lobe points, indicated with the red dot in figure~\ref{fig:verification}. The field strength here is 0.4759, resulting in a very similar area.

The last verification is to directly calculate $\oint_{\rm lobe} \AAA\cdot\mathrm{d}l$ over the boundary of the lobe by a simple Riemannian sum over the segments of the lobes.
The four different calculations show good correspondence, with the crude triangle approximation of course being the least accurate with an error of about 4\%, with the Biot-Savart evaluation on the boundary having a relative difference of about $5\cdot10^{-4}$ and the shoelace area formula a relative difference of $10^{-5}$.

\subsection{QUASR  229079}
\begin{figure*}[htbp]
  \begin{overpic}[width=\linewidth]{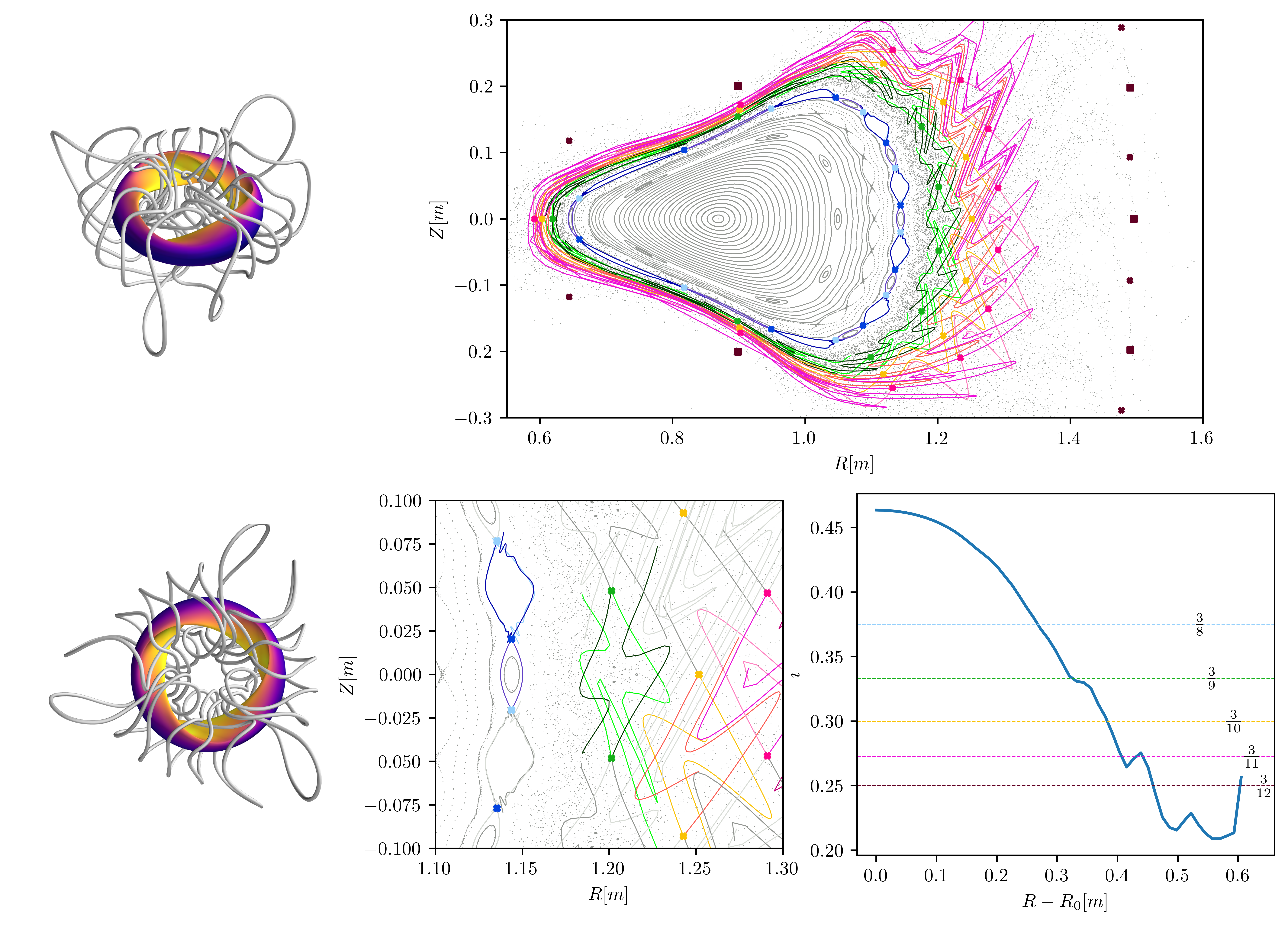}
    \put(3, 68){\textbf{(a)} }
    \put(3, 32){\textbf{(b)} }
    \put(40, 68){\textbf{(c)} }
    \put(35, 31){\textbf{(d)} }
    \put(77, 31){\textbf{(e)} }
  \end{overpic}
  \caption{Analysis of QUASR configuration 229097.
  (a), (b): Coils and a magnetic surface of the configuration colored by the magnetic field strength.
  (c): Poincar\'e section at $\phi=0$ plane, showing an ordered center surrounded by a large chaotic region that contains many high-order resonances.
  The manifolds associated with the 6/16 (light blue/blue and violet/purple), 3/9 (light-/dark green), 3/10 (yellow/orange) , and 3/11 (pink/fuchsia) are shown, as well as the fixed points of a 3/12 chain in pink.
  (d): zoom in of the interacting turnstiles in the chaotic edge, with only the manifolds around a single island resonance zone colored.
  (e): rotational transform profile calculated along the midplane.
  \label{fig:quasr_manyisland}}
\end{figure*}
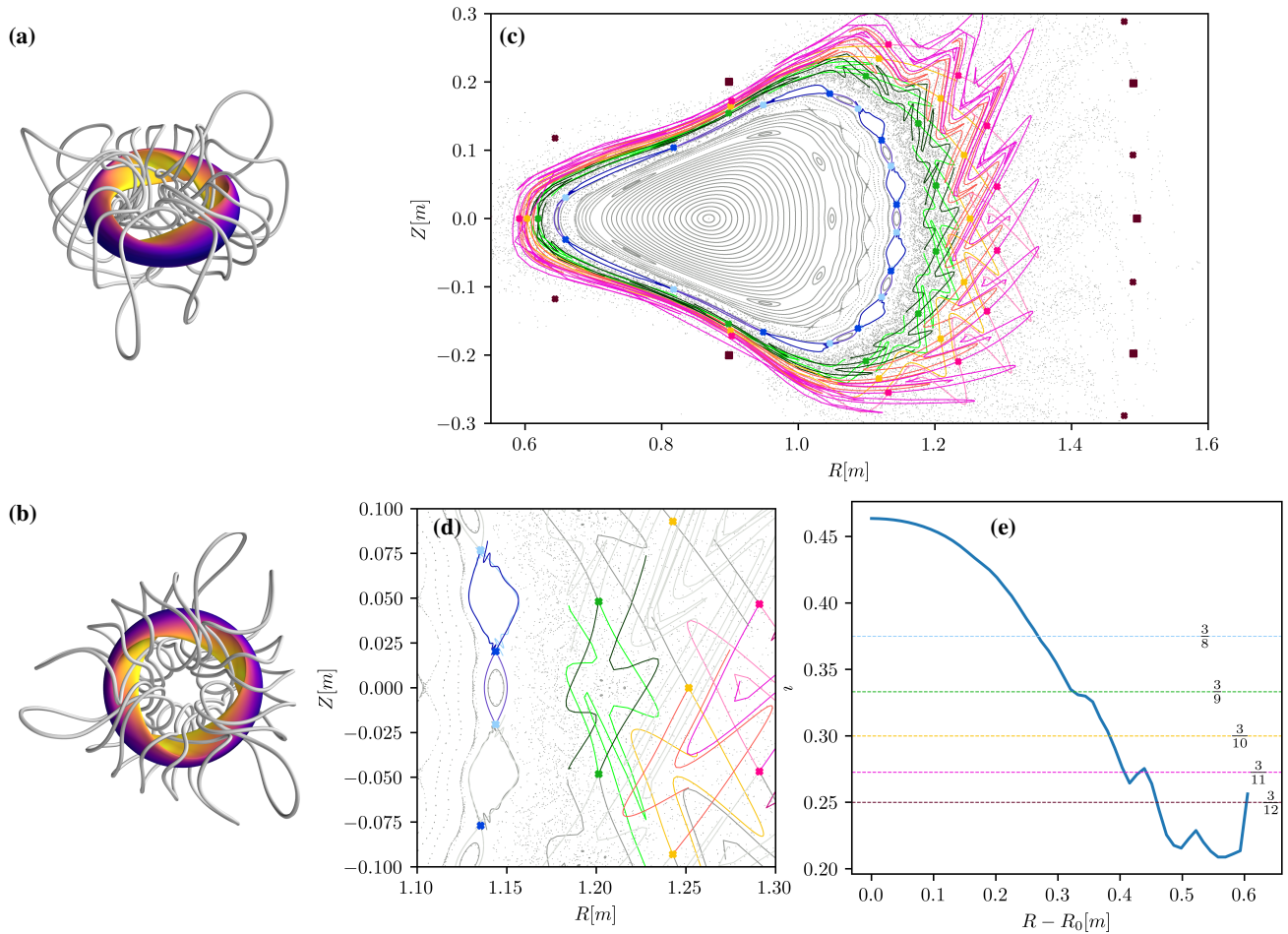

We next analyze the structures in the edge of QUASR configuration 229079. This is a three field-period quasi axisymmetric stellarator with 18 coils and a low rotational transform.
The coils and a flux surface of this configuration are shown in figure~\ref{fig:quasr_manyisland} (a)-(b).
This configuration is rather compact, and has strongly shaped ridges on the inboard side of the tori which create rotational transform, very similar to the configurations generated by~\citet{plunk2020perturbing} through a perturbative modification of axisymmetric equilibria.
Configurations such as this, though not explicitly optimized for, are common in the QUASR database among the high aspect-ratio low rotational transform configurations.
\citet{Henneberg2024compact} have investigated the potential of such configurations as a reactor dubbed the 'Compact Stellarator-tokamak hybrid'.

In the Poincar\'e plot of this configuration we see closed nested surfaces in the center, with an occasional island, but further out the field becomes more and more chaotic. 
The innermost island chain for which we calculate manifolds, is a period-doubled 6/16 island chain, shown in blues and purples. 
After that there is little discernible structure in the Poincar\'e plot. 
Nevertheless, there are many fixed points in this region. 

In this chaotic region, we can still calculate the approximate rotational transform profile. We do this by only integrating for 20 field periods, and finding the average angle a field line rotates around the axis for this number of mappings. This is shown in figure~\ref{fig:quasr_manyisland} (e). 
We also plot the rational numbers corresponding with specific rationals. 

Even though the field outside the 6/16 resonance is completely chaotic, we can still find hyperbolic fixed points associated with the crossed resonances, and we can calculate their manifolds and turnstile areas. 
Figure~\ref{fig:quasr_manyisland} (d) shows a zoom in to these manifolds, with only one occurrence of each colored. 
We see the turnstile area increase significantly with respect to the size of the resonance zone from each resonance zone to the next.
This is quantified in table~\ref{tab:quasr_manyisland_table}.
We again observe that the turnstile area is larger on the outboard side of the configuration. 

We can see that in the first island chains are clearly distinguishable, and separate, with a small turnstile area relative to the resonance zone area. 
In such a region, individual island chains act as partial barriers, and field lines must thread through two small turnstiles, both on the inboard and on the outboard side of the chain to pass through. 
As one moves further and further into the chaotic region, the manifolds of subsequent island chains start to align, they become large with respect to the resonance zone area. 
In this case there is a large exchange of flux between one side of the island and the other. 
We also see that the turnstile on one periodic orbit has a large overlap with the turnstile on the next periodic orbit. 
In this regime, the resonances do not act as barriers at all, and field lines can pass from one resonance zone to the next in a few mappings. 
We speculate that the onset of turnstile overlap between neighboring resonances can cause a change in the character of the transport.

It is also interesting to note that the rotational transform profile seems to increase towards the very edge of the region where the field line map can be calculated, and crosses the 3/12 resonance a second time. 
This indicates that there are nontwist map dynamics~\cite{del1996area} at play in this outermost chaotic region.
We can even locate the periodic orbit associated with the 3/12 resonance in this very edge of the configuration, but because this region is close to the coils, many of the field lines here do not make a full orbit, and the manifolds cannot be calculated. 

\begin{table}
\begin{tabular}{lcc}
\hline
Manifold & Turnstile Area (in) & Turnstile Area (out) \\
\hline
manifolds 6/15 a& $5.8090\cdot 10^{-07}$ & $2.5502\cdot 10^{-06}$ \\
manifolds 6/15 b& $5.2863\cdot 10^{-10}$ & $4.4831\cdot 10^{-10}$ \\
manifolds 3/9 & $3.6184\cdot 10^{-05}$ & $7.3541\cdot 10^{-05}$ \\
manifolds 3/10 & $1.3430\cdot 10^{-04}$ & $2.1494\cdot 10^{-04}$ \\
manifolds 3/11 & $2.5547\cdot 10^{-04}$ & $3.7709\cdot 10^{-04}$ \\
\hline
\end{tabular}

  \caption{Turnstile areas of select resonances in the QUASR 229079 configuration. }\label{tab:quasr_manyisland_table}
\end{table}

    \subsection{\label{sec:W7-X} Turnstile area in the chaotic low-iota configurations in the W7-X stellarator}
Finally we will use the tools to calculate the turnstile area in configurations of the Wendelstein 7-X (W7-X) stellarator.
W7-X was optimized for neoclassical transport and low bootstrap current.
It is therefore said to be a quasi-isodynamic stellarator.
W7-X has already shattered the records for stellarator field accuracy~\cite{pedersen2016confirmation}, stellarator fusion triple product~\cite{dinklage2018magnetic}, and stellarator confinement~\cite{beidler2021demonstration}.

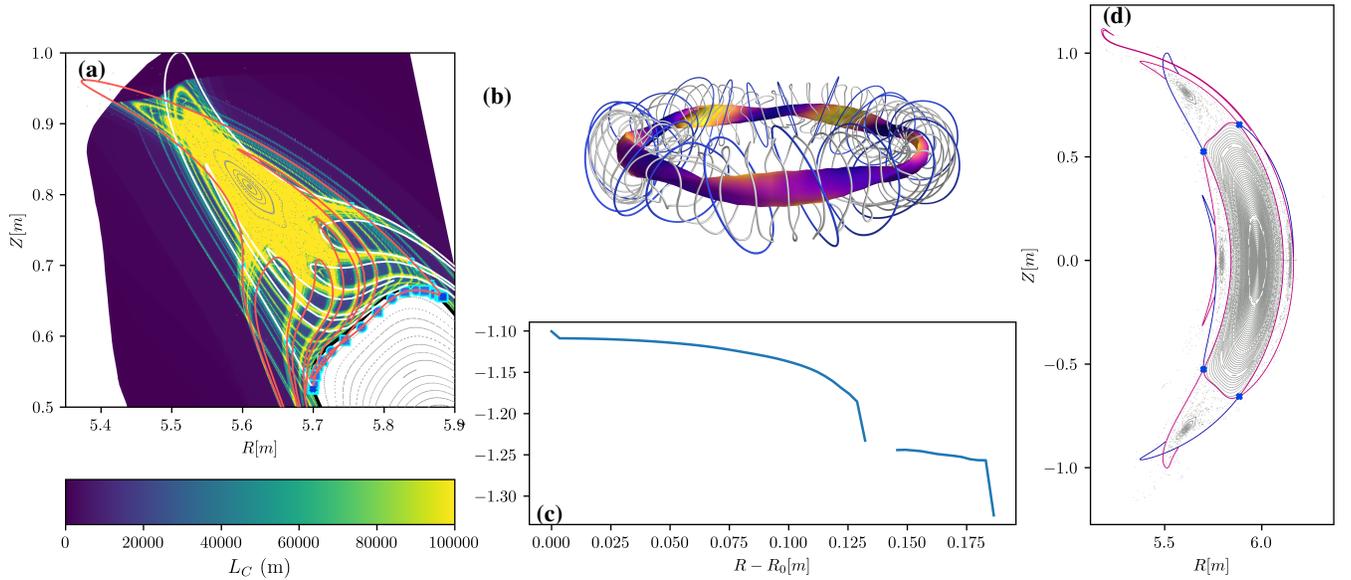
\begin{figure*}[htbp]
  \begin{overpic}[width=\linewidth]{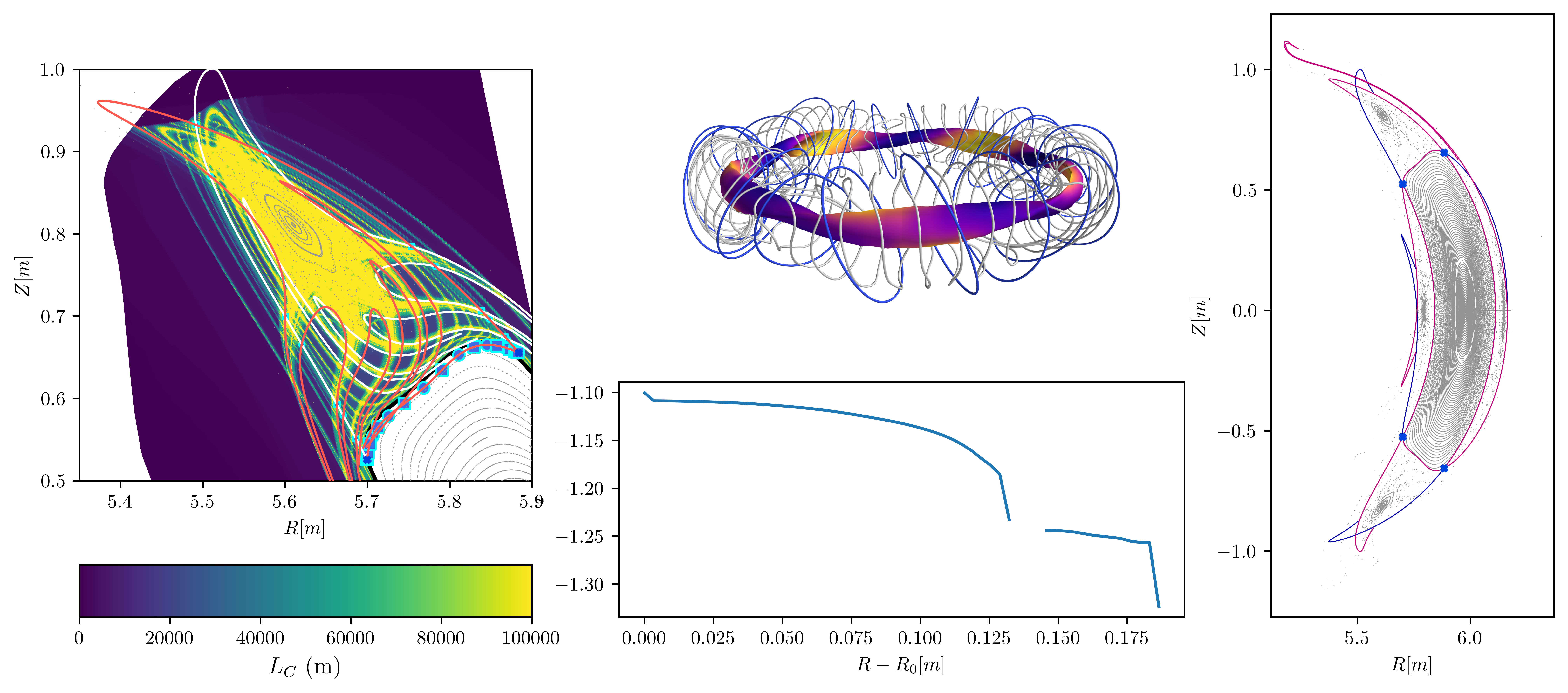}
    \put(6, 38){\textbf{(a)} }
    \put(36, 36){\textbf{(b)} }
    \put(40, 5){\textbf{(c)} }
    \put(82, 42){\textbf{(d)} }
  \end{overpic}
  \caption{The W7-X GYM00+1750 configuration.
  (a): Connection length plot showing the minimum distance (both forwards and backwards along field lines) to the PFCs. Overlaid are the unstable (stable) manifolds of the 5/4 island chain in white (orange).
  The features in the connection length plot correlate strongly with the manifolds of the outer heteroclinic connection.
  The inner heteroclinic connection has an small turnstile area compared to the outer heteroclinic connection.
  (b): Modular coils (white) and planar coils (blue) of the W7-X stellarator, as well as a magnetic surface, colored by the magnetic field strength on it.
  (c): Rotational transform in the GYM configuration.
  (d): Poincar\'e section of the full $\phi=0$ cross section, showing nested surfaces inside and a very chaotic edge.
    \label{fig:w7x}}
\end{figure*}

W7-X is a five field period device that creates the magnetic field with a total of 70 main coils\footnote{there are also five windowpane coils and ten trim coils that can be used to further adjust the field, but do not generate the bulk of the confinement field} that generate the field.
Per half-field period, there are 5 unique \emph{modular} coils, that have complicated shapes and two \emph{planar} coils.
The coil configuration and the shape for the plasma region is shown in figure~\ref{fig:w7x} (b).
These are copied around the device according to stellarator symmetry and field periodicity to produce the full coil set.
The power supplies to these superconducting coils are connected according to this symmetry, but the current in each of these 7 unique coil groups can be set separately.

When the modular coils carry their design current of 1.62 MA each, and there is no current in the planar coils, the stellarator generates the \emph{standard configuration}.
This configuration has a large volume of closed flux surfaces, that is terminated with a large 5/5 island chain that acts as a divertor and determines the strikepoints of the plasma on the PFCs (and the location of heat loading).

One of the design goals for W7-X  was also its versatility as a research reactor, and different configurations can be achieved by changing the currents in the coils.
The strongly shaped modular coils create the intricately shaped and twisted magnetic field, whereas the modular coils are so oriented as to provide a field along the axis without any additional shaping.
By running a current to the planar coils which opposes the toroidal field, an increase of twist is achieved resulting in a configuration with higher rotational transform.

When the current in the modular coils is set to $I_{\rm modular} = 1.1095MA$, and the current in the planar coils is $I_{\rm planar} = -\tfrac{1}{3} I_{\rm modular} \simeq -0.3661MA$, we get a configuration where the rotational transform in the edge is $\imath =1.25$.
A Poincar\'e section of this field is shown in figure~\ref{fig:w7x} (d), and the resultant rotational transform profile is shown in figure~\ref{fig:w7x} (c).  
This configuration is called GYM00-1750 and is studied in the context of nonresonant divertors.
One of the interesting features of this configuration is that the edge of the field is chaotic, which has led to its study in the context of nonresonant divertors~\cite{davies2025topological, boozer2018simulation}.

One method that is often employed to design divertors is to analyze the connection length in the divertor region, with tools such as EMC3-Lite.
The EMC3-Lite code is capable of calculating heat loads on plasma-facing components (PFCs) and PFC-to-PFC connection lengths for a given magnetic field and PFC arrangement \cite{feng2022review}.
For both, it relies on following Monte Carlo particles (which either simply follow magnetic field lines or take a random walk to imitate anisotropic diffusion), which are tracked on a field-aligned grid (for more details on the grid scheme see \cite{feng2005simple}).
The PFC-to-PFC connection length $L_c$ is specific to each field line and is not recorded, but the grid cell-averaged connection length $<L_c>$ is calculated and recorded, and is useful for revealing the structure of the magnetic field and predicting regions of heat loading on PFCs (see for example \cite{feng2022review, gao2023improvement}).
$<L_c>$ is calculated by tracing Monte Carlo particles both forwards and backwards along the field until they intersect a PFC. The connection length of this field line contributes to all grid cells it passes through, and $<L_c>$ for a given cell is the arithmetic mean of $L_c$ of all unique field lines passing through the cell.
The cell in which each magnetic field line is initialized is selected to ensure that each cell contains at least $n_\text{min}$ unique trajectories. The location of an initialised field line in a cell is uniformly random.
An upper cutoff $L_{c,\text{max}}$ is applied to each field line, since field lines which do not intersect a PFC (intact surfaces around island O-points, for example) would otherwise be traced indefinitely.
In the result presented here  we select $n_\text{min}=100$ and $L_{c,\text{max}}=10^4$ meters. The grid resolution used here is 480 points in the radial direction and 1025 points in the poloidal direction. This is several times larger than is typically required for heat deposition calculations but is used here to show the fine structure emerging from the magnetic chaos.

The connection length in the island region is plotted in figure~\ref{fig:w7x} (a), and shows a remarkable pattern of alternating long connection length and short connection length regions.
The manifolds of the outer heteroclinic connection are also shown on this plot, and there is a strong correlation between the two.
We attribute the small discrepancies between the two do differences in the integrator and coil representation. 

The turnstile mechanism helps to explain how this pattern comes about: the region that is not in the turnstile, but is close to the island, will not be mapped out of the island resonance zone in the first many mappings, and has a large connection length. 
One needs to be in a turnstile to have a low $<L_c>$, but since it measures PFC to PFC connection length, only being in one turnstile is not sufficient.
We therefore see a lattice of alternating long and short connection length, caused by trajectories that are in both turnstiles alternating with trajectories that are only in one, in an ever repeating pattern that Poincar\'e dubbed a 'trellis'~\cite{poincare1893methodes}, made visible in our analysis of the stellarator divertor!

Just like the QUASR configurations, we again see that the turnstile area on the outer heteroclinic connection is much larger than the turnstile on the inner heteroclinic connection.
The oscillations of the inner manifold are not visible at all at this scale of magnification.
Heat and particles that cross the inner separatrix is channeled along the stable (unstable) manifold, where the above-mentioned trellis structure determines which trajectories eventually intersect the wall.
The chaos originating from the turnstile on this outer heteroclinic connection is interesting to study, and is likely to affect the locations where the plasma intersects the wall, but the rate of transport into this region due to chaos in the field is very low due to the small inner turnstile area. 

We have seen in all the stellarator configurations, that on each resonance the turnstile on the outer heteroclinic connection is significantly larger than the turnstile on the inner heteroclinic connection.
This is likely because the perturbations that induce the chaos are in part due to ripple effects caused by the finite number of discrete coils.
The field lines of the outer turnstile pass much closer to these coils, and are more strongly affected by this ripple.

In terms of stellarator divertor design, it would be interesting to understand if this feature is unavoidable, or just a feature of the configurations analyzed.
In other words, would it be possible to generate a field with coils, where the turnstile on the inner separatrix is large, potentially larger than that of the outer separatrix?
This would enhance the transport from the plasma into the resonance zone at a controllable rate, and would be interesting to contrast with configurations like GYM00+1750 where the chaos is mostly caused by dynamics originating on the outer manifolds.

\section{\label{sec:conclusions} Conclusions and Discussion}
We have illustrated how to use fixed points of the magnetic field line map to analyze chaos and transport in fusion reactor magnetic fields.
For this, we have adapted methods used to analyze transport in Hamiltonian dynamical systems to the specific case of analyzing magnetic connectivity in the edge of fusion reactors, by exploiting the mathematical equivalence between $1\tfrac{1}{2}$ dimensional Hamiltonian systems and magnetic field line flow in three dimensions.
Specifically we have implemented a calculation by~\citet{meiss2015thirty} based on the action principle to calculate the \emph{turnstile area}, a direct measure for chaotic transport across transport barriers formed by homo- and heteroclinic connections.
Such structures are intentionally created in the edge of fusion reactors to control the outflow of heat from the reactor to the PFCs.
We have demonstrated this calculation in the two-dimensional iterated map (the Tokamap), in perturbed axisymmetric fields similar to the tokamak fusion reactor, in fully three-dimensional stellarator fields from the QUASR database and in chaotic configurations of the W7-X fusion reactor.

We have adapted and optimized the calculation for the specific application to three-dimensional magnetic fields, where the integration of field lines is a relatively costly operation.
With the analytical toy model we have shown how different perturbations generate chaotic transport and increase the turnstile area.
We have also shown that this calculation is capable of resolving the turnstile area in a very large range of perturbation amplitudes, from very nearly integrable fields, to almost completely chaotic fields.

Transport through an island chain requires a trajectory (field line) to pass through two different turnstiles one after another.
The inner turnstile governs passage from the plasma region into the resonance zone, and the outer turnstile then determines passage into the outer chaotic region.
If, as in most devices, including the W7-X stellarator, the PFCs are placed intersecting an edge island, the most important transport channel is likely to be through the inner turnstile, as in the island region many trajectories quickly intersect a PFC.
The calculations of turnstile area in the stellarator configurations show that the outer turnstile area is much larger than the inner turnstile area in the observed configurations.
This is likely due to the perturbations due to coil ripple, and the outer manifold passing closer to the coils.
Though such fields may look chaotic, the transport from the plasma to the PFCs is hardly affected by chaos originating from the outer turnstile.

In a fusion reactor, field line flow is not the only factor determining heat transport.
It is true that heat conductivity along field lines is several orders of magnitude higher than perpendicular conductivity, but there are other processes that compete.
In stellarators there can be significant neoclassical transport~\cite{beidler2021demonstration}, caused by the fact that particle orbits do not follow field lines perfectly.
In a stellarator field a significant fraction of particles can lie on orbits that have a net radial displacement, and the net outflow of high-energy particles from the core can cause significant heat transport.
Modern optimized stellarators control this by generating fields that minimize such particle orbits, and this has been proven to work in the neoclassically optimized stellarator W7-X~\cite{beidler2021demonstration}.
Recently there has also been significant interest in quasisymmetric optimized stellarators, and configurations have been found where nearly all particles are confined~\cite{landreman2022magnetic}.

In an optimized stellarator, where neoclassical transport is very low, turbulent transport becomes the dominant mechanism that transports heat perpendicular to the field lines.
Turbulence can be caused by many mechanisms, and a predictive theory of turbulent transport in stellarators is nowhere near.
Transport caused by the turnstile mechanism is therefore not always the most important factor determining heat transport in the stellarator edge.
Especially if the turnstile area is small, it is to be expected that other mechanisms such as neoclassical or turbulent transport can dominate.
The calculation of the turnstile area allows for direct quantification of the amount of transport that is due to chaos in the magnetic field, and would allow one to potentially separate these two transport channels in experiment.

It is important to note that particle trajectories are not identical to field line flow, and especially for high-energy particles the particle trajectories can be misaligned with the magnetic surfaces~\cite{chambliss2024fast}.
Particle motion in three dimensions is governed by its' own three-dimensional Hamiltonian system (with six-dimensional phase space).
The dimensionality can be reduced by neglecting the gyromotion (the rapid precession of particles perpendicular to the field), and identifying a spatial coordinate as 'time-like', to yield a four-dimensional Hamiltonian.
\citet{chambliss2024fast} brilliantly use this to provide insight into energetic particle transport in stellarators by sectioning this system at specific pitch angles, such that a 1$\tfrac{1}{2}$-D system is achieved which chaos, islands and resonances can be observed in their two-dimensional sections.
It would be interesting to apply turnstile calculations to these reduced-dimensional maps to quantify the rates of energetic particle transport in these sections.
Furthermore, transport could be analyzed in the higher-dimensional Hamiltonians without restricting to sections by evaluating the equivalent metrics in the four- or higher dimensional systems.

The calculation of the turnstile area can be applied to a wide range of circumstances, but it still requires user intervention to set up.
The fixed points can be found, but choosing the manifolds to trace along and the amount of mappings to use is a process that varies a lot from configuration to configuration.
Other approaches, such as level-set learning~\cite{ruth2025level} could provide a faster and less geometry-specific methods to approach chaos minimization.

The turnstile area also correlates with the chaos in the field, and when the turnstile area is zero, the stable and unstable manifold must overlap.
The calculation of the turnstile area can be a valuable tool for designing future stellarators, especially when the dynamics on a specific heteroclinic connection (i.e. the inboard or the outboard side of the island) is important for the transport.
Using such calculations, the amount of chaos in the edge region can be tuned, both on the inner and on the outer turnstile of an island chain.
The capability to optimize the turnstile area is in the process of being developed and will be the subject of future publications.

\begin{acknowledgments}
  The authors would like to thank Jim Meiss for the inspiring discussions that led to this work about implementing the action calculation in magnetic fields.
  The authors would like to thank Matt Landreman for invaluable feedback in general, and in particular his suggestions for improving the heteroclinic trajectory finding algorithm.
  Similarly, our eternal gratitude goes out to Andrew Giuliani for creating and maintaining the QUASR database, which provides a near endless supply of interesting configurations to analyze.
R. Davies would like to acknowledge useful guidance from and discussions with Oliver Schmitz, Kelly Garcia and Dieter Boeyaert (University of Wisconsin-Madison) and Sophia Henneberg, Sergei Makarov and  Amit Kharwandikar (Max Planck Institute for Plasma Physics, Greifswald.
This work was supported by a grant from the Simons Foundation\ (1013657, JL). 
This work has been carried out within the framework of the EUROfusion Consortium, partially funded by the European Union via the Euratom Research and Training Programme (Grant Agreement No 101052200 — EUROfusion). The Swiss contribution to this work has been funded by the Swiss State Secretariat for Education, Research and Innovation (SERI). Views and opinions expressed are however those of the author(s) only and do not necessarily reflect those of the European Union, the European Commission or SERI. Neither the European Union nor the European Commission nor SERI can be held responsible for them.
\end{acknowledgments}

\bibliography{references}

\appendix
\section{vector potential of the toy model field}

The toroidal magnetic field $B^\phi$ in equation~\eqref{eq:bphi} is written in terms of a vector potential.
Using gauge freedom, the vector potential is written as a single function $A^R$.
This vector potential is found using mathematica and has the form:
\begin{widetext}
\begin{align*}
    A^R = &\frac{1}{4r} \text{Real}[ \left( 4 \cdot q_a + s \cdot \left( 5r^2 - 10rR + 4R^2 + 2(z - Z)^2 \right) \right) \cdot \sqrt{-r^2 + 2rR - (z - Z)^2} \cdot (z - Z)\\
    &- i r(r - 2R) \cdot \left( 4 \cdot q_a + \left( 3r^2 - 6rR + 4R^2 \right) \cdot s \right) \cdot \log\left( -i z + \sqrt{-r(r - 2R) - (z - Z)^2} + i Z \right)].
\end{align*}
\end{widetext}

\end{document}